\documentclass[aps,prl,twocolumn,floatfix,superscriptaddress,preprintnumbers,nofootinbib]{revtex4-1}
\usepackage{graphicx}
\usepackage{epsfig}
\usepackage{amsmath}
\usepackage{amsfonts}
\usepackage{float}
\usepackage{amssymb}
\usepackage{color}
\usepackage[bottom]{footmisc}
\usepackage{array,multirow,makecell}
\usepackage{slashed}
\usepackage[colorlinks,citecolor=blue]{hyperref}
\usepackage{ulem}
\usepackage[utf8]{inputenc}
\usepackage[english]{babel}
\usepackage{url}
\usepackage{hyperref}
\usepackage{xcolor}
\usepackage{subfigure}
\usepackage{lipsum}
\usepackage[bottom]{footmisc}
\usepackage{pifont}
\begin{document}
\title{A Solar Investigation of Multicomponent Dark Matter}

\author{Amit Dutta Banik}
\email[E-mail: ]{amitdbanik@gmail.com}
\affiliation{Physics and Applied Mathematics Unit, Indian Statistical Institute, Kolkata-700108, India}


\begin{abstract}

If multiple thermal weakly interacting massive particle (WIMP) dark matter candidates exist, then their capture and annihilation dynamics inside a massive stars such as Sun could change from conventional method of study. With a simple correction to time evolution of dark matter (DM) number abundance inside the Sun for multiple dark matter candidates, significant changes in DM annihilation flux depending on annihilation, direct detection cross-section, internal conversion and their contribution to relic abundance are reported in present work.   
\end{abstract}


\maketitle

\section {I. Introduction}

Various astrophysical observations clearly indicates the existence of dark matter in the Universe. Experiments like Planck \cite{Aghanim:2018eyx} also claim existence of dark matter that constitutes about 80\% matter content of the Universe itself. Despite clear astrophysical and gravitational evidences, basic constituent of dark matter still remains a riddle. Among many of proposed DM candidates, weakly interacting massive particle (WIMP) remains a promising candidate for dark matter, discussed in many literatures. Different experiments are searching for existence of WIMP like dark matter via direct and indirect method. Direct search experiments such as XENON1T \cite{Aprile:2018dbl,Aprile:2015uzo}, XENONnT \cite{XENON:2020kmp,XENON:2023sxq}, PandaX-II \cite{Cui:2017nnn}, PICO \cite{PICO:2019vsc} etc. search for dark matter scattering off target nucleus inside detector, and provide limits on DM spin-independent and spin-dependent scattering cross-sections. On the other hand galactic centre, dwarf galaxies are naturally dark matter abundant as they could capture DM particles gravitationally. These captured DM, can also undergo annihilation into different Standard Model (SM) particles and generate gamma-ray, positron etc. are treated as source of DM indirect detection. Indirect detection experiments like Fermi-LAT \cite{Fermi-LAT:2015att}, DES \cite{Fermi-LAT:2015ycq}, MAGIC \cite{MAGIC:2016xys}, H.E.S.S. \cite{HESS:2016mib} provides upper limits on dark matter annihilation cross-sections based on the search of gamma-rays originating from galactic centre and dwarf galaxies. Similarly excess  positron or proton from DM annihilation are probed by AMS-02 \cite{AMS:2014bun,AMS:2018avs}, H.E.S.S. \cite{Aharonian:2008aa}, Fermi-LAT \cite{Abdollahi:2017nat} and DAMPE \cite{TheDAMPE:2017dtc}. Apart from these, dark matter can also be captured by astrophysical objects like Sun and then annihilate into neutrinos, thus act as a source for DM indirect detection. The produced neutrinos from DM annihilation, after leaving Sun, reaches earth based neutrino detectors through various internal process such as oscillation, absorption etc. 
If capture and annihilation of dark matter reaches steady state inside Sun, neutrino detectors like IceCube \cite{IceCube:2016dgk}, Super-K \cite{Super-Kamiokande:2015xms}, ANTARES \cite{ANTARES:2016xuh} at earth can provide bounds on the DM-nucleon scattering cross-section (both spin-dependent and spin-independent). 
Different studies have been performed in this context to develop the theory for dark matter indirect detection signature from Sun and explore many interesting aspects such as iso-spin violating nature of dark matter, self scattering of dark matter, nature of DM anihilation flux \cite{Faulkner:1985rm,Griest:1986yu,Gould:1987ju,Gould:1991hx,Jungman:1995df,Bertone:2004pz,Barger:2007xf,Hooper:2008cf,Belotsky:2008vh,Wikstrom:2009kw,Erkoca:2009by,Zentner:2009is,Covi:2009xn,Chen:2011vda,Bernal:2012qh,Chen:2014oaa,Catena:2016ckl,Tiwari:2018gxz,Gaidau:2018yws,Gupta:2022lws,Bose:2022ola}. 
However, if steady state is not achieved inside Sun, neutrino detectors could only provide us with information of dark matter annihilation flux (or muon flux) observed by the detector. This could happen if the DM-nucleon scattering cross-section is very small. Interestingly, such a situation may also occur if one consider there exists multiple dark matter candidates with annihilation into hidden sector which changes the dynamics of DM capture and annihilation. The concept of multi-particle dark matter is very intriguing with basic assumption of existence of more than one dark matter candidate. Different phenomenological models for multicomponent WIMP dark matter have already been pursued in many literatures \cite{Feldman:2010wy,Bhattacharya:2013hva,Bian:2013wna,Esch:2014jpa,Bhattacharya:2016ysw,Ahmed:2017dbb,Herrero-Garcia:2017vrl,Aoki:2018gjf,Herrero-Garcia:2018qnz,Chakraborti:2018aae,Elahi:2019jeo,Borah:2019aeq,Bhattacharya:2019fgs,Barman:2018esi,Bhattacharya:2019tqq,DuttaBanik:2020jrj,DiazSaez:2021pfw}. However, very few attempts were made to study multicomponent DM annihilation signatures from Sun but eventually simplified either into single component DM annihilation with some specific condition or some other studies involving monochromatic neutrino signal from Sun \cite{Aoki:2012ub,Aoki:2014lha,Berger:2014sqa,Aoki:2017eqn}, and thus, features of multi-particle DM has not been addressed in details. 
In this work, we investigate how dark matter indirect detection from massive stars like Sun could change in presence of multiple WIMPs that not only annihilate in SM but also undergo annihilation into themselves. The present works reports how annihilation flux of DM candidates could change significantly due to multicomponent nature and internal conversion. For a simple two component scenario, it is shown that flux of one dark matter candidate could get suppressed due to annihilation into another candidate while the flux of other DM candidate gets boosted significantly. It is also found that the changes in DM annihilation flux also depends on
relic abundance of each dark matter candidate, strength of their DM-nucleon scattering cross-section and annihilation cross-sections significantly. In the present work, we explore the consequence of such effects in detail for a two component WIMP dark matter formalism. The paper is organised as follows. In Sec.~II, after a small note on the dynamics of single component DM inside Sun with
relevant bounds, the formalism of multicomponent dark matter evolution inside Sun is developed and the results obtained depending on various parameters are presented. In Sec.~III, the results from multicomponent dark matter study are briefly summarised and some future prospects are addressed with concluding remarks.

\section{II. Multicomponent Dark Matter at Sun}
\label{Sun}
The formalism of multi-particle dark matter capture and annihilation inside massive stars such as Sun is developed in this section.
To begin with, a small summary of standard conventional DM capture and annihilation inside the Sun is presented for single component dark matter and later concept of multi-particle dark matter scenario is addressed.
\subsection{Standard Scenario} 
Dark mater, commonly considered as WIMP particles, undergoes scattering inside a stellar body like Sun and gets captured
if their velocity falls below the escape velocity of Sun. 
Captured DM particles can also annihilate into Standard Model particles. For a dark matter candidate $\chi$ of mass $m_{\chi}$, with nucleon proton scattering cross-section $\sigma_{\chi p}$ and annihilation cross-section $\langle \sigma v \rangle$, the evolution of DM number abundance inside Sun is given as     
\begin{align}
\frac{dN}{dt} = C_{\text{c}}  -C_{a}  N^2.
\label{1}
\end{align}
In the Eq.~(\ref{1}) above, $N$ denotes number of DM in the Sun, $C_{\text{c}}$ and $C_{a}$ are the capture rate and annihilation rate coefficient of dark matter. 
Dark matter capture rate when only spin-independent (SI)  scattering is given as~\cite{Jungman:1995df,Bertone:2004pz}
\begin{align}
C_{\text{c}}\simeq1.24\times10^{24}\textrm{ s}^{-1}\left(\frac{\rho_{0}}{0.3\textrm{ GeV/cm}^{3}}\right)\left(\frac{270\textrm{ km/s}}{\bar{v}}\right)^{3} \nonumber \\ \left(\frac{\textrm{GeV}}{m_{\chi}}\right)^{2}
\left(\frac{2.6\sigma_{\textrm{H}}^{\textrm{SI}}+0.175\sigma_{\textrm{He}}^{\textrm{SI}}}{10^{-6}\textrm{ pb}}\right).
\label{eq:capture_SI}
\end{align}
where 
\begin{equation}
\sigma_{i}^{\textrm{SI}}=A^{2}\left(\frac{m_{A}}{m_{p}}\right)^{2}\left(\frac{m_{\chi}+m_{p}}{m_{\chi}+m_{A}}\right)^{2}\sigma_{\chi p}^{\textrm{SI}}\, .
\label{SI}
\end{equation}
In Eq.~(\ref{SI}), only scattering of DM with  Hydrogen and Helium is considered as massive stars like Sun are mostly comprised of these two light elements.
The annihilation rate  coefficient of dark matter inside Sun is parametrised by $C_{a}$ \cite{Griest:1986yu},
\begin{equation}
C_{a}\simeq\frac{\left\langle \sigma v \right\rangle V_{2}}{V_{1}^{2}},
\label{Cann}
\end{equation}
with 
\begin{equation}
V_{j}\simeq6.5\times10^{28}\textrm{ cm}^{3}\left(\frac{10\textrm{ GeV}}{jm_{\chi}}\right)^{3/2}.
\label{vol}
\end{equation}
Using Eqs.~(\ref{eq:capture_SI})-(\ref{vol}) a straight forward solution to Eq.~(\ref{1}) can be obtained of the form
\begin{align}
N(t) = \sqrt{\frac{C_{c}}{C_{a}} }\tanh \left( \frac{t}{\tau_{\odot}} \right),
\label{eqn:1DM}
\end{align}
for initial condition $N(t=0)=0$ where  $\tau_{\odot}^{-1} \equiv\sqrt{C_{\text{c}} C_{a}}$
is the time scale for equilibrium condition when $\frac{dN_{\chi}}{dt}=0$ is reached. 
The time to reach steady state is determined by scattering and annihilation cross-section of dark matter candidate
and may be larger than the age of Sun. However, if $\tau_{\odot}$ is smaller than age of Sun, then DM annihilation inside Sun
can be expressed in terms of capture rate $\Gamma_{\rm{ann}}=C_c/2$ and number of dark matter at steady state is $N_{EQ}=\sqrt{\frac{C_c}{C_a}}$. 
Similarly if one considers capture of dark matter is driven by spin-dependent (SD) dark matter nucleon scattering, the capture rate is then expressed as \cite{Jungman:1995df,Bertone:2004pz}
\begin{align}
C_{\text{c}}\simeq3.35\times10^{24}\textrm{ s}^{-1}\left(\frac{\rho_{0}}{0.3\textrm{ GeV/cm}^{3}}\right)\left(\frac{270\textrm{ km/s}}{\bar{v}}\right)^{3} \nonumber \\
\left(\frac{\textrm{GeV}}{m_{\chi}}\right)^{2}\left(\frac{\sigma_{\textrm{H}}^{\textrm{SD}}}{10^{-6}\textrm{ pb}}\right).
\label{eq:capture_SD}
\end{align}
with the DM-nucleus scattering cross-section
\begin{equation}
\sigma_{i}^{\textrm{SD}}=A^{2}\left(\frac{m_{\chi}+m_{p}}{m_{\chi}+m_{A}}\right)^{2}
\frac{4(J_i+1)}{3 J_i}\left|\left\langle S_{p,i}\right\rangle +\left\langle S_{n,i}\right\rangle \right|^{2}
\sigma_{\chi p}^{\textrm{SD}}
\end{equation}
where $\left\langle S_{p,i}\right\rangle$ ($\left\langle S_{n,i}\right\rangle$) denotes expectation value of proton (neutron) averaged over all nucleons of the nucleus. As mentioned before, if the dark matter capture and annihilation inside Sun does not reach steady state,
then it is convenient to use DM annihilation rate as $\Gamma_{\rm{ann}}=\frac{C_a}{2} N(t_S)^2$ where $t_S=4.6\times10^9$ year. Therefore, the quantity $\Phi$, known as dark matter annihilation flux can be written as
\begin{align}
\Phi = \frac{\Gamma_{\text{ann}}}{4\pi D^2}\, ,
\label{phi}
\end{align}
where $D$ is the distance from the source (Sun) to observer at earth. 

We solve for number of dark matter accumulated in Sun and then calculate the flux $\Phi$ for both spin-independent and spin-dependent scattering of dark matter.
For this purpose, we use stringent limits on DM-nucleon SI scattering cross-section obtained from XENONnT \cite{XENON:2020kmp}, limit on SD scattering cross-section from PICO \cite{PICO:2019vsc} and IceCube \cite{IceCube:2016dgk} experiment assuming thermal annihilation of WIMP dark matter $\langle \sigma v \rangle \simeq 2.2\times 10^{-26}$ cm$^3$ s$^{-1}$ \cite{Steigman:2012nb}
\footnote{We have used the projected sensitivity of DM-nucleon scattering cross-section \cite{XENON:2020kmp} instead of the recent upper limit from XENONnT \cite{XENON:2023sxq} as bound obtained from Ref.~\cite{XENON:2020kmp} is more stringent.}. Therefore, one can obtain the DM annihilation flux using $\Gamma_{\rm{ann}}=\frac{C_a}{2} N(t_S)^2$ and Eq.~(\ref{phi}).
It is to be noted that WIMP annihilation cross-section $\langle \sigma v \rangle$ at present day might be small due to velocity or momentum suppression which depends on the nature of interaction. Therefore, only the specific scenarios are considered where DM annihilation cross-section 
is not suppressed and comparable to thermal annihilation cross-section $\langle \sigma v \rangle \simeq 2.2\times 10^{-26}$ cm$^3$ s$^{-1}$ \cite{Steigman:2012nb}\footnote{For example their are possible pseudo scalar mediated interaction of dark matter $\bar{\chi}\gamma^5\chi\phi$ or axial vector interaction via spin-1 boson $\bar{\chi}\gamma^{\mu}\gamma^5\chi V_{\mu}$ that allows low velocity
$\langle \sigma v \rangle$ without suppression.}.

Before we move on to further studies with multiple dark matter candidates, let us now try to emphasise in brief what would happen 
if we have a dark matter candidate partially contributing to DM relic abundance
which can undergo spin-independent or spin-dependent scattering and annihilation
into SM sector only. Consider a dark matter candidate that constitutes a fraction $f$ of total dark matter abundance $f=\frac{\Omega_{\chi}h^2}{\Omega_{\rm{DM}}^{\rm Tot}h^2}$, where $\Omega_{\rm{DM}}^{\rm Tot}h^2=0.1199\pm 0.0027$ as observed by Planck \cite{Aghanim:2018eyx}. This would modify the number evolution of dark matter inside Sun as
\begin{align}
\frac{dN}{dt} =f C_{\text{c}}  -\frac{C_{a}}{f}  N^2\, ,
\label{1a}
\end{align}
where we $f$ actually scales the scattering cross-section $\sigma_i^{\prime}=f\sigma_i$ and DM annihilation cross-section
$\langle \sigma v \rangle^{\prime}=\frac{\langle \sigma v \rangle}{f}$, with $\langle \sigma v \rangle \simeq 2.2\times 10^{-26}$ cm$^3$ s$^{-1}$, arising due to partial contribution to total DM abundance as $\Omega_{\rm DM}h^2\propto \frac{1}{\left\langle \sigma v \right\rangle}$. It is to be noted that observation of gamma-ray flux in the galactic centre and dwarf galaxies by various indirect search experiment provide limits on the DM annihilation cross-section. In the present work, for a dark matter candidate with relic abundance 10\% of total DM abundance $\Omega_{\rm{DM}}^{\rm Tot}h^2$, we restrict ourself to the region of parameter space $200~{\rm{GeV}}< m_{\chi} \leq 1000$ GeV, consistent with limits from combined study by MAGIC and Fermi-LAT \cite{MAGIC:2016xys} (for DM annihilation into $\mu^+\mu^-$ channel) and H.E.S.S. \cite{HESS:2016mib} (for DM annihilation into $W^+W^-$ channel using Einasto 2 profile) experiment.

\begin{figure}
    \begin{center}
        \includegraphics[width=0.45\textwidth]{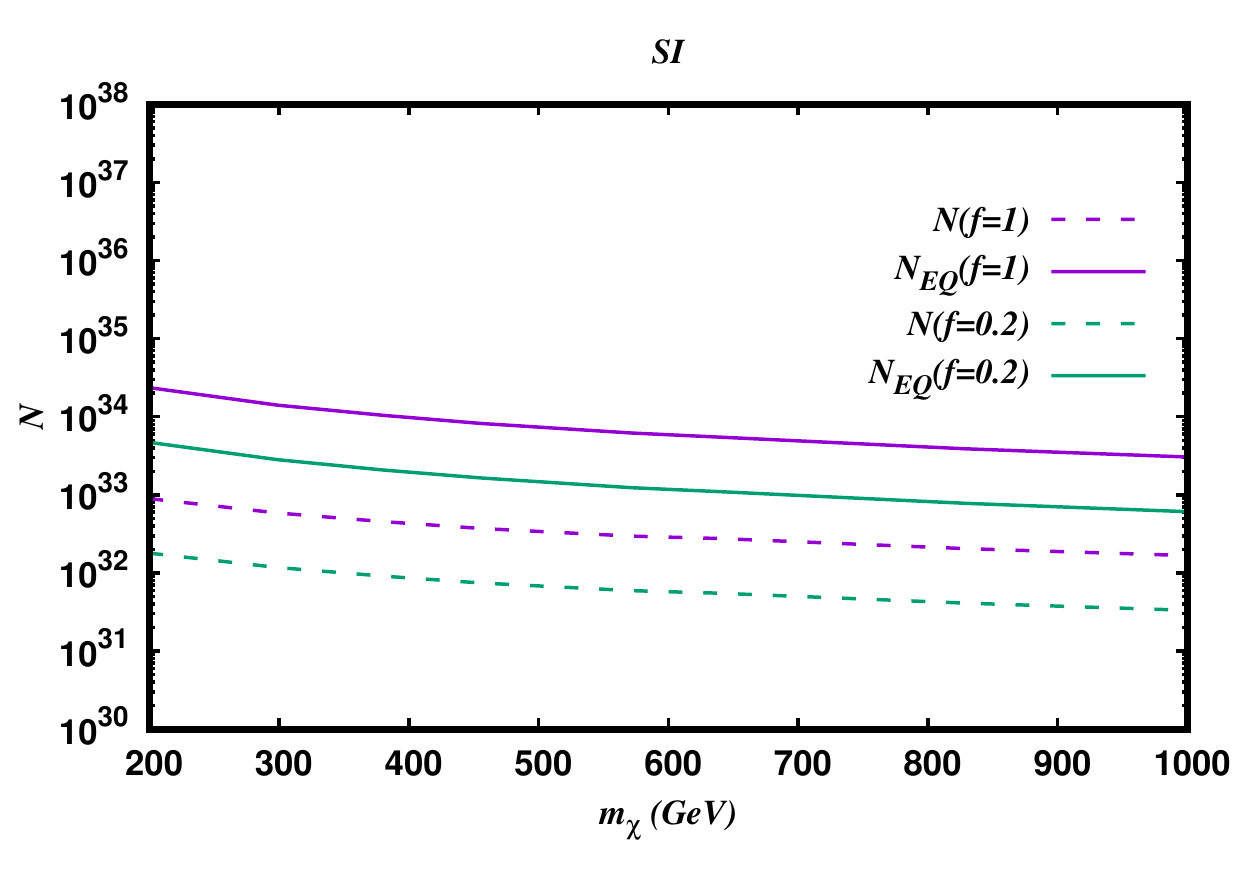}
        \includegraphics[width=0.45\textwidth]{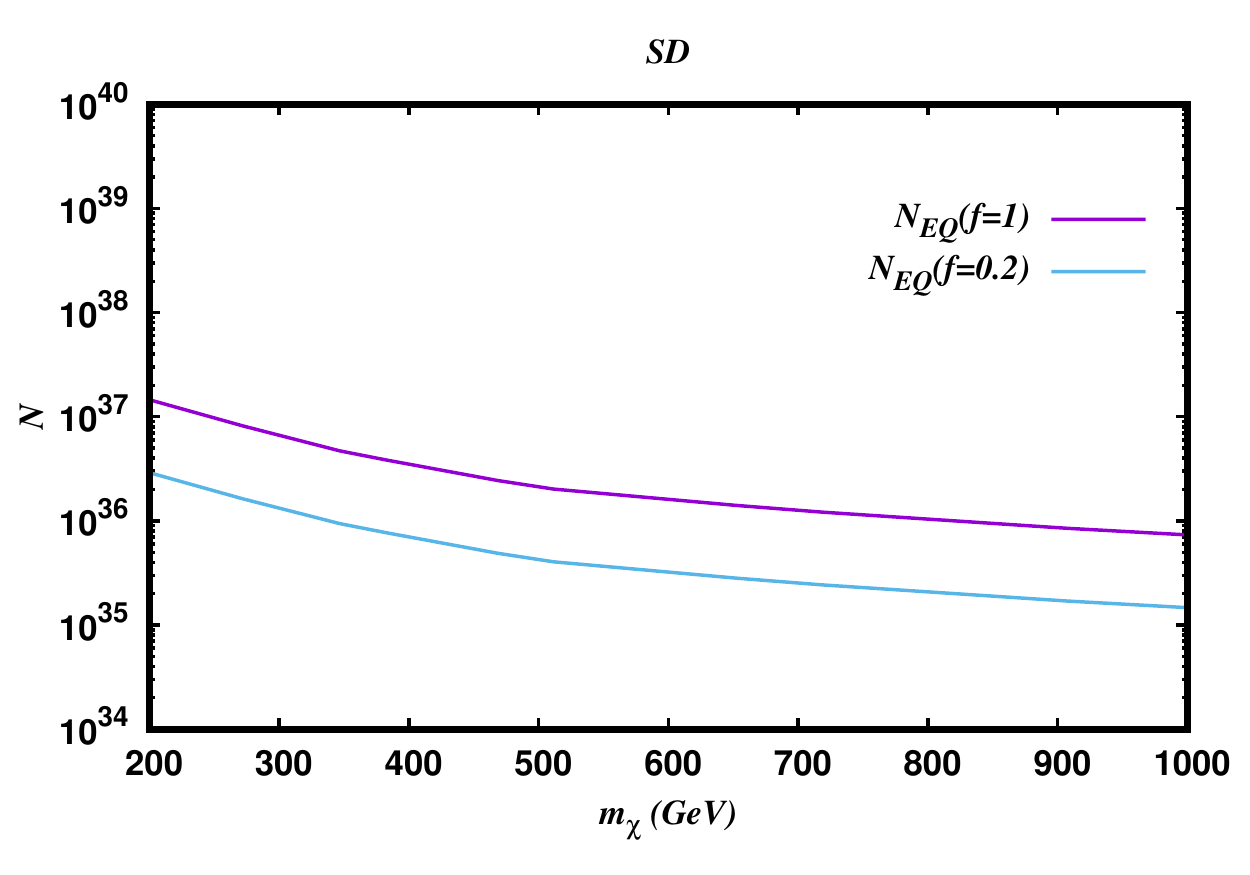}
            \caption{\it Upper panel: Number abundance of dark matter inside Sun for SI scattering of dark matter at $t=t_S$ (dashed lines) and at $t=\tau_{\odot}>t_S$ (solid lines). Lower panel: Number abundance (equilibrium) of dark matter inside Sun for SD scattering of dark matter at $t=t_S>\tau_{\odot}$. Variation of number abundance is observed for two values of $f=1$ and $f=0.2$, $f$ being the fraction of total DM relic abundance.}
        \label{fig:a}
    \end{center}
\end{figure}

\begin{figure}
    \begin{center}
        \includegraphics[width=0.45\textwidth]{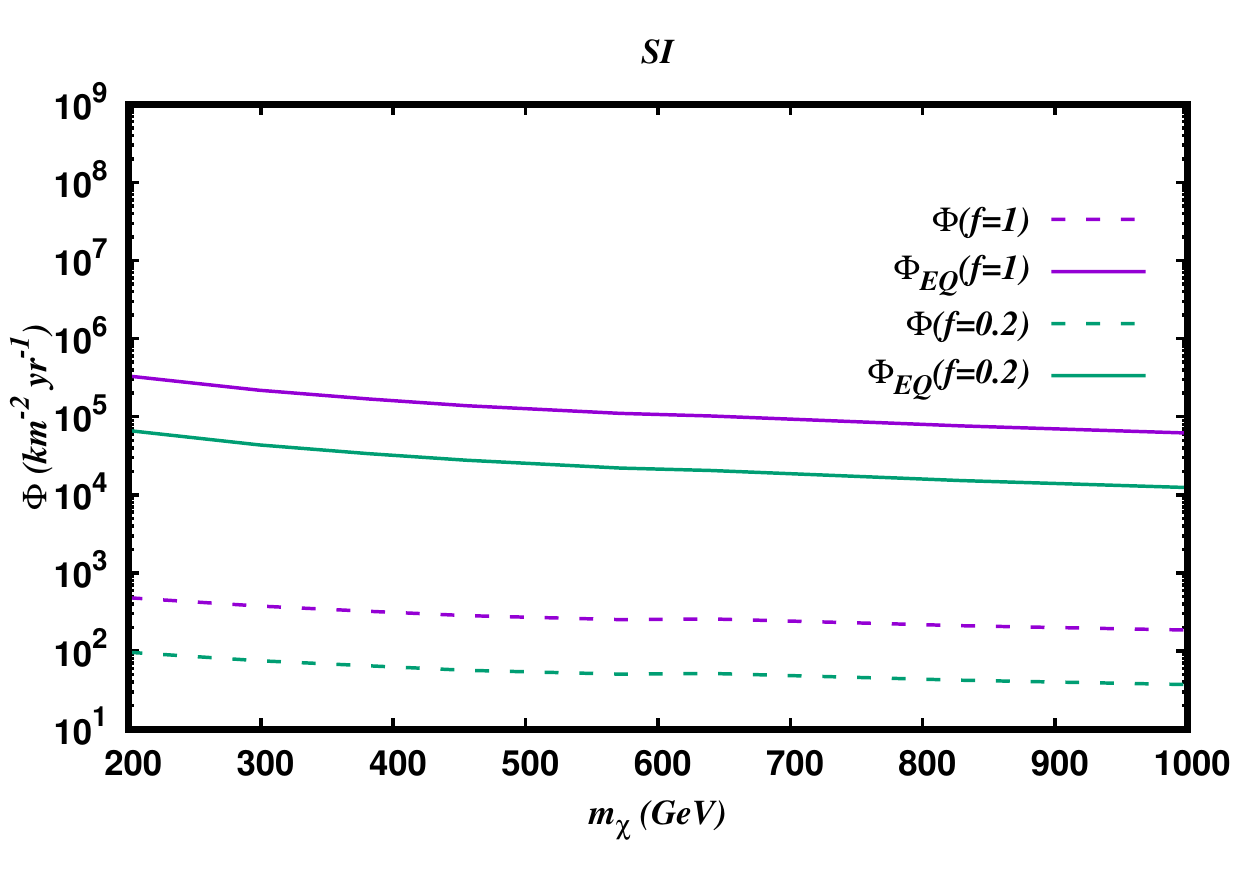}
        \includegraphics[width=0.45\textwidth]{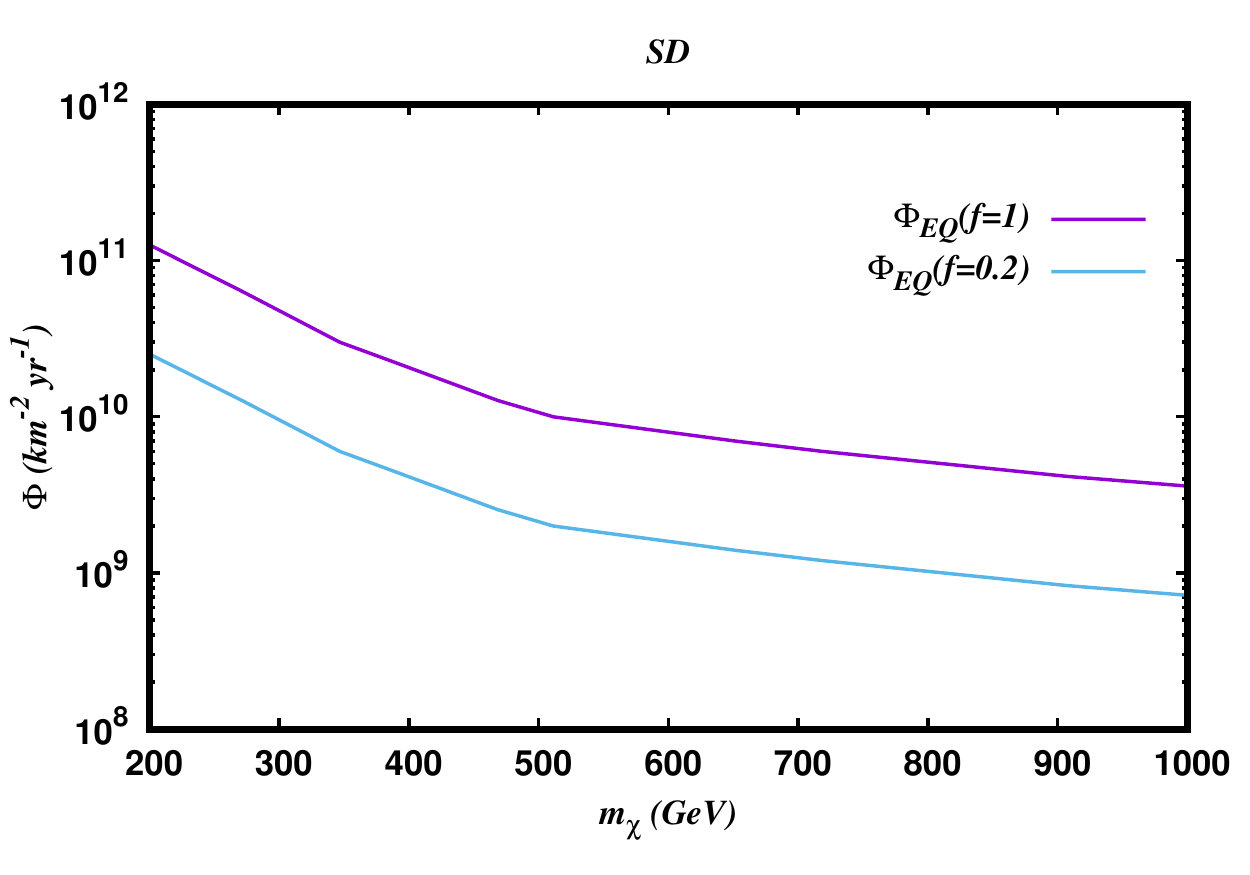}
            \caption{\it Upper panel: Dark matter annihilation flux inside Sun for SI scattering of dark matter at $t=t_S$ (dashed lines) and at equilibrium $t=\tau_{\odot}>t_S$ (solid lines). Lower panel: Dark matter annihilation flux inside Sun for SD scattering of dark matter at $t=t_S>\tau_{\odot}$ in equilibrium. Both figures are plotted for $f=1$and $f=0.2$.}
        \label{fig:b}
    \end{center}
\end{figure}
 
Solution to Eq.~(\ref{1a}) is simple and one finds it to be similar to Eq.~(\ref{eqn:1DM}) with $N(t) = f\sqrt{\frac{C_{c}}{C_{a}} }\tanh \left( \frac{t}{\tau_{\odot}} \right)$ and equilibrium abundance $N_{EQ}= f \sqrt{\frac{C_c}{C_a}}$. It is interesting to mention that,
the time required to reach steady state remains unaltered even for dark matter with fractional abundance for ${\tau^{\prime}_{\odot}} \equiv \frac{1}{\sqrt{fC_{\text{c}} \frac{C_{a}}{f}}}={\tau_{\odot}}$.
Flux of dark matter annihilation $\Phi^{\prime} =\frac{C_{a}}{2f} N^2=f\Phi$ is also scaled by factor $f$ for a given dark matter candidate with fractional DM abundance.   
In Fig.~\ref{fig:a}, we plot the dark matter number abundance for SI and SD DM-nucleon scattering inside the Sun using bounds on DM-nucleon SI and SD scattering for different dark matter mass with  $f=1$ and $f=0.2$. 
For the spin-independent DM-nucleon scattering with most sensitive limit from XENONnT \cite{XENON:2020kmp}, we observe dark matter number abundance fails to reach equilibrium number density as $t=t_S$. Depletion in number abundance is observed when fractional contribution of dark matter is considered. 
However, for spin-dependent interaction, DM number abundance reaches steady state due to large scattering cross-section and this holds even for fractional contribution as illustrated in lower panel of Fig.~\ref{fig:a}. Similar to the case of SI DM scattering, here we also observe depletion in DM number abundance for partial DM abundance scaled by the factor $f$. With the obtained number abundance of dark matter, in Fig.~\ref{fig:b} we plot the upper limit on DM annihilation flux $\Phi$ in km$^{-2}$ yr$^{-1}$ for SI and SD DM-nucleon scattering for different dark matter mass. Due to stringent constraints on DM spin-independent scattering, DM annihilation flux is found to be very much suppressed $({\cal{O}} \simeq 10^6$) when compared with flux achieved from DM spin-dependent scattering. For partial contribution of dark matter candidate, the DM annihilation flux is further reduced by factor $f$.
Therefore, if dark matter partially contributes to the total DM relic abundance, we expect reduction in maximum DM flux produced in general in absence of conversion between dark sector. Solving Eq.~(\ref{1}) and Eq.(\ref{1a}), we observe DM annihilation flux has simple correlation $\Phi^{\prime}=f\Phi$. It is to be noted that, DM annihilation flux is of great importance as differential neutrino flux from Sun at earth (at the detector) is directly related to the quantity $\Phi$, expressed as \cite{Baratella:2013fya}
\begin{equation}
\frac{d\Phi_{\nu_i}}{dE_{\nu_i}}=\Phi\left(\frac{dN_{\nu_i}}{dE_{\nu_i}}\right)_{x}\, ,
\label{eq:neutrino_flux}
\end{equation}
where $\frac{dN_{\nu_i}}{dE_{\nu_i}}$ is the energy spectrum spectrum of $\nu_i$ and $\bar{\nu_i}$ ($i=e,\mu,\tau$) produced per annihilation taking into account all possible effects of medium (hadronization, neutrino absorption, hadron stopping etc) and neutrino oscillation for DM annihilation into specific final state $x$ ($x=$ quark, lepton or gauge boson).
Hence, differential neutrino flux for fractional DM contribution is given as
\begin{equation}
\frac{d\Phi^{\prime}_{\nu_i}}{dE_{\nu_i}}=\Phi^{\prime}\left(\frac{dN_{\nu_i}}{dE_{\nu_i}}\right)_{x}=f\frac{d\Phi_{\nu_i}}{dE_{\nu_i}}\, .
\label{eq:neutrinoflux}
\end{equation}
Therefore, neutrino flux at the detector is also scaled by factor $f$ when dark matter candidate 
shares $f$ fraction of total DM relic abundance.

The standard mechanism to study the neutrino flux from dark matter is to perceive up-going muons into the detector
generated by muon neutrinos interacting with the detector target material to avoid irreducible background effects from down going muons. The muon flux 
originating from annihilation of dark matter with mass $m_{\chi}$ can be expressed as \cite{Hooper:2008cf,Erkoca:2009by,Covi:2009xn,Chen:2011vda} 
\begin{eqnarray}
\Phi_{\mu} &=& \int_{E_\mu^{\rm th}}^{m_\chi} d E_\mu \int_{E_\mu}^{m_\chi} d E_{\nu_\mu} \frac{d \Phi_{\nu_\mu}}{d E_{\nu_\mu}}   \times\nonumber \\ 
&& \left[ \frac{\rho}{m_p} \frac{d \sigma_\nu}{d E_\mu} (E_\mu, E_{\nu_\mu}) R_\mu (E_\mu, E_\mu^{\rm th}) \right] + (\nu\to \bar{\nu})\, , \nonumber \\  
\label{fluxmu}
\end{eqnarray}
where $\rho$ denotes density of water or rock. The expression for weak interaction neutrino (anti neutrino) scattering cross-section
is given as \cite{Barger:2007xf,Barger:2001ur}
\begin{eqnarray}
\frac{d \sigma^{(p,n)}_\nu(E_\mu, E_{\nu_\mu})}{d E_\mu}  = \frac{2}{\pi} G_F^2 m_p \left( a_\nu^{(p,n)} + b_\nu^{(p,n)} \frac{E_\mu^2}{E_{\nu_\mu}^2}\right)\,, \nonumber \\
\end{eqnarray}
where $a_\nu^{(p,n)}=0.15,~0.25$, $b_\nu^{(p,n)}=0.04,~0.06$ for neutrinos while for antineutrinos $a_{\bar \nu}^{(p,n)}=b_\nu^{(n,p)}$,
 $b_{\bar \nu}^{(p,n)}=a_\nu^{(n,p)}$.
The range of length up to which muon travels before losing their energy below threshold energy $E_\mu^{\rm th}$ of the detector is expressed as
\begin{eqnarray}
R_\mu (E_\mu, E_\mu^{\rm th}) = \frac{1}{\beta \rho} \log \left( \frac{\alpha + \beta E_\mu}{\alpha + \beta E_\mu^{\rm th}} \right)\,,
\end{eqnarray}
where $\alpha=2.3\times 10^{-3}\,{\rm cm}^2{\rm g}^{-1} {\rm GeV}^{-1}$ 
and $\beta = 4.4\times 10^{-6}\,{\rm cm}^2{\rm g}^{-1}$. The muon flux $\Phi_{\mu}$ obtained from muon neutrinos generates muon events inside the detector depending on the exposure time and effective area (or volume) of detector, which is treated as signal for DM detection. Null detection of any such excess event provides upper limit on the flux $\Phi_\mu$. 

From Eqs.~(\ref{eq:neutrino_flux})-(\ref{fluxmu}), we can clearly state that if a dark matter $\chi$ is a component with relic abundance $f\Omega_{\rm{DM}}^{\rm Tot}h^2$, the muon flux generated at the detector is simply given as $\Phi_{\mu}^{\prime}=f\Phi_{\mu}$ where $\Phi_{\mu}$ given by Eq.~(\ref{fluxmu}). 
However, for multi-particle dark matter, number evolution of dark matter candidates can be coupled via exchange interaction, which is a key signature of multiple dark matter scenario. Thus, we now move further towards a complete picture of multi-particle DM annihilation in the next subsection.

\subsection{Multicomponent approach} 
Let us now derive the evolution equations for multiple dark matter candidates inside Sun. To explore with the minimal scenario, a two component
dark matter evolution inside the Sun is assumed. Let us consider two WIMP dark matter candidates $\chi_1$ and $\chi_2$ of mass $m_1$ and $m_2$ which contribute to total relic abundance of dark matter $\Omega_{\rm{DM}}h^2=\Omega_{1}h^2 + \Omega_{2}h^2=f\Omega_{\rm{DM}}^{\rm Tot}h^2+(1-f)\Omega_{\rm{DM}}^{\rm Tot}h^2$. The fractional contribution of first DM candidate $\chi_1$ in DM abundance is $f$ and that of the second DM candidate is $1-f$.
In this context, we consider two cases to be studied\newline
 I) interaction between $\chi_1$ and $\chi_2$ is absent or negligible and
 \newline 
 II) $\chi_1$ and $\chi_2$ has significant interaction that induces DM conversion $\chi_1 \chi_1 \leftrightarrow \chi_2 \chi_2$. \\
 \subsubsection{Case I}
 Without putting much effort, one can easily derive equations for number evolution of DM candidates with no mutual interaction, are given as
\begin{align}
&\frac{dN_1}{dt} = fC_{1c}  -\frac{C_{1a}}{f}  N_1^2\, ,\nonumber \\
&\frac{dN_2}{dt} = (1-f)C_{2c}  -\frac{C_{2a}}{(1-f)}  N_2^2\, ,
\label{2}
\end{align} 
where $N_{1,2}$ denotes number of $\chi_{1,2}$ with initial condition $N_{1,2}(t=0)=0$ for both candidates. Parameters $C_{kc}$
and $C_{ka}$; $k=1,2$ are defined as capture rate and annihilation rate coefficients of DM candidates similar to Eq.~(\ref{eq:capture_SI})-(\ref{vol}) with $C_{kc} \propto \sigma_{\chi_{k} p}$ and $C_{ka} =\left\langle \sigma v \right\rangle V_{2a}/V_{1a}^2$ respectively where we assume $\langle \sigma v \rangle \simeq 2.2\times 10^{-26}$ cm$^3$ s$^{-1}$. The factor $f$ or $(1-f)$ thus suppress the DM-nucleon scattering and enhances the DM annihilation cross-section. 

Solutions to number evolution of DM candidates are similar to Eq.~(\ref{eqn:1DM}), expressed as
\begin{align}
& N_{1}(t) = f \sqrt{\frac{C_{1c}}{C_{1a}} }\tanh \left( \frac{t}{\tau_{1\odot}} \right)\, ,\nonumber \\ 
& N_{2}(t) = (1-f) \sqrt{\frac{C_{2c}}{C_{2a}} }\tanh \left( \frac{t}{\tau_{2\odot}} \right)\, ,
\label{eqn:2DM}
\end{align}
where $\tau_{k\odot}=\frac{1}{\sqrt{C_{kc} C_{ka}}};~k=1,2$, is independent of $f$.
This situation, is not very much interesting as it appears to be same as single dark matter component case. 
Therefore, we now focus on to the other proposed scenario where dark matter candidates interact with each other. 
 \subsubsection{Case II}
Let us now discuss the scenario where evolution of dark matter particles are considered including hidden sector annihilation.
For two dark matter candidates it simply states that the annihilation between dark sector particles are significant resulting the conversion
$\chi_1 \chi_1 \leftrightarrow \chi_2 \chi_2$ to dominate or contribute in the evolution equations mentioned in Eq.~(\ref{2}).
Let us now consider the case $m_1>m_2$ resulting hidden sector annihilation of heavier DM candidate $\left\langle \sigma v \right\rangle_{11\rightarrow 22}$. This will modify the evolution equations which can be written as follows
\begin{align}
&\frac{dN^{\prime}_1}{dt} = fC^{\prime}_{1c}  -C^{\prime}_{1a}  N_1^{\prime 2}-C^{\prime}_{12}  N_1^{\prime 2}\, , \nonumber \\
&\frac{dN^{\prime}_2}{dt} = (1-f)C^{\prime}_{2c}  -C^{\prime}_{2a}  N_2^{\prime 2}+C^{\prime}_{12}  N_1^{\prime 2}\, ,
\label{3}
\end{align} 
where coefficients $C^{\prime}_{ka},~k=1,2$ and $C^{\prime}_{12}$ are denoted as
\begin{equation}
C^{\prime}_{ka}= \left\langle \sigma v \right\rangle_{kk}\frac{V_{2k}}{V_{1k}^{2}}\, , \hskip 5mm 
C^{\prime}_{12} = \left\langle \sigma v \right\rangle_{11\rightarrow 22}\left[\frac{V_{2}}{V_{1}^{2}}\right]_{m=m_1}\, .
\label{C12}
\end{equation}
Without any loss of generality, one can assume $C^{\prime}_{kc}=C_{kc}$ for $k$-th dark matter candidate with fixed mass
$m_k$ and scattering cross-section $\sigma_{\chi_k p}$. The reverse process $\chi_2 \chi_2 \rightarrow \chi_1 \chi_1$ is not taken into account as dark matter is considered non-relativistic inside the Sun. 
It is to be noted that since dark matter candidates annihilate into each other, the simple correlation
$\Omega_{\rm DM}h^2\propto \frac{1}{\left\langle \sigma v \right\rangle}$ between dark matter annihilation 
 and its contribution to total DM relic abundance is lost. Indeed, this also leads to solution of coupled Boltzmann equations to obtain 
 relic abundances of DM candidates, which have been studied in many literatures, discussed briefly in Appendix A. Therefore, the parameter for fractional contribution $f$, will not be able to decide corresponding annihilation cross-sections to be precise as appeared before where conversion process between dark sector is absent. As a result annihilation cross-sections into SM sector $\left\langle \sigma v \right\rangle_{kk}$ and annihilation $\left\langle \sigma v \right\rangle_{11\rightarrow 22}$ are now treated as free parameters of equations. 
The solution to $N^{\prime}_1(t)$ will be of similar form as described in Eq.~(\ref{eqn:2DM}) and one obtains
\begin{align}
& N_{1}^{\prime}(t) = \sqrt{\frac{fC^{\prime}_{1c}}{C^{\prime}_{1a}+C^{\prime}_{12}} }\tanh \left( \frac{t}{\tau^{\prime}_{1\odot}} \right)\, ,\nonumber \\
&\tau^{\prime}_{1\odot}=\frac{1}{\sqrt{fC^{\prime}_{1c}(C^{\prime}_{1a}+C^{\prime}_{12})}\, ,
}\,
\label{N1}
\end{align}
where $C^{\prime}_{1c}=C_{1c}$ for fixed mass $m_1$ and $\sigma_{\chi_1 p}$. Interestingly, time to reach steady state inside Sun now depends on $f$
as observed from Eq.~(\ref{N1}) rather than being independent of $f$. 
However, solution to $N^{\prime}_2(t)$ will be different as it is coupled to the solution of $N^{\prime}_1(t)$ of Eq.~(\ref{3}).
Let us try to look for an approximate solution to $N^{\prime}_2(t)$, assuming that $\chi_1$ dark matter has reached its steady sate. In this situation the term $C^{\prime}_{12}  N_1^{\prime 2}$ is constant and for $C^{\prime}_{12}  N_1^{\prime 2}>>(1-f)C^{\prime}_{2c}$, one can easily obtain a solution
\begin{align}
& N_{2}^{\prime}(t) \simeq \sqrt{\frac{C^{\prime}_{12}}{C^{\prime}_{2a}}}\sqrt{\frac{fC^{\prime}_{1c}}{C^{\prime}_{1a}+C^{\prime}_{12}} }\tanh \left( \frac{t}{\tau^{\prime}_{2\odot}} \right)\, ,\nonumber \\
&\tau^{\prime}_{2\odot} \simeq \frac{1}{\sqrt{C^{\prime}_{2a}C^{\prime}_{12}}}\sqrt{\frac{C^{\prime}_{1a}+C^{\prime}_{12}}{fC^{\prime}_{1c}}\, .
}\,
\label{N2}
\end{align}
Eventually it can be observed that $N_{2}^{\prime}$ is much larger than $N_{2}$ leading to enhancement in $\chi_2$ abundance inside Sun whereas
$\tau^{\prime}_{2\odot}<<\tau_{2\odot}$, and thus $\chi_2$ reaches equilibrium earlier than expected.
Interestingly, both $N_{2}^{\prime}$ and $\tau^{\prime}_{2\odot}$ are now dependent on the evolution parameters of $\chi_1$.
For better accuracy, one needs to solve for Eq.~(\ref{3}) numerically to obtain number of dark matter candidates accumulated inside Sun at
$t=t_S$, denoted as $N_{1,2}^{\prime}(t_S)$. Both scattering and annihilation of dark matter depends on the nature of dark matter candidates. As we have mentioned, if DM annihilates into SM sector only, then $\langle \sigma v \rangle$ determines its contribution to  total DM relic abundance. Therefore, using $\langle \sigma v \rangle \simeq 2.2\times 10^{-25}$ cm$^3$ s$^{-1}$ for $f\geq0.1$, we consider $200~{\rm{GeV}}<m_1\leq 1000$ GeV and $m_1>m_2$ in present analysis consistent with indirect detection bounds from gamma-ray observations following Refs. \cite{MAGIC:2016xys,HESS:2016mib} mentioned eralier. With conversion process $\chi_1 \chi_1 \rightarrow \chi_2 \chi_2$ in effect Eq.~(\ref{3}) is solved and compared with the case when hidden sector DM annihilation is absent (Case I). 

The study of single component dark matter reveals that DM annihilation flux obtained from spin-independent dark matter is negligible compared to that obtained for spin-dependent dark matter as direct detection experiments impose stringent bound on DM-nucleon spin-independent scattering cross-section compared to spin-dependent dark matter scattering cross-section.
This is due to the fact that spin-independent DM scattering depends on the mass of scattering nucleus and direct detection experiments with heavy nuclei has better sensitivity to SI interactions. 
On contrary, indirect detection by neutrino detector experiments are sensitive to spin-dependent dark matter detection with respect to spin-independent DM and provides more stringent bound than direct detection experiments for $m_{DM}\geq 100$ GeV as observed by IceCube \cite{IceCube:2016dgk}.
For DM annihilation inside Sun, with rich abundance of Hydrogen makes it a very good probe towards SD dark matter interaction and indirect detection by neutrino detectors.
It is naturally assumed that dark matter accumulated inside massive stars like Sun are completely thermalised which is primary requirement for the present study. Thus DM thermalise quickly upon capture or thermalisation time is much smaller compared to the age of Sun such that most of the DM are thermalised. Hence, for multicomponent scenario, we consider spin-dependent dark matter interactions
having mass range $200~{\rm{GeV}}<m_1 \leq 1000$ GeV (indirect detection limits from gamma-ray flux with $f\geq0.1$) and $m_1>m_2$ with a large allowed range of DM-nucleon scattering cross-section region $\sigma_{\chi_k p}^{\rm{SD}}= 10^{-48}-10^{-41}$ cm$^2$ where DM is thermalised. 
For simplicity for the rest of the study we denote spin-dependent DM-nucleon cross-section as $\sigma_{\chi_k p}^{\rm{SD}}=\sigma_{\chi_k p}$ as the analysis hereafter only considers SD dark matter scattering.

Let us briefly describe a feasible particle physics scenario where such multicomponent dynamics with two spin-dependent dark matter candidates can be visualised. Consider two fermionic dark matter candidates $\chi_1$ and $\chi_2$ both having axial vector type interactions $\chi_k\gamma_\mu \gamma_5 \chi_k;~k=1,2$ that interact with SM sector fermion bilinear  $\bar{f}_{\rm SM}\gamma_\mu \gamma_5 f_{\rm SM}$ via some vector boson mediator. The term $\chi_1\gamma_\mu \gamma_5 \chi_2$ is forbidden with simple assumption that $\chi_1$ and $\chi_2$ are charged under different $Z_2$ symmetry.
This will allow both dark matter candidates to have spin-dependent scattering leaving a large range of allowed DM-nucelon scattering parameter space while interaction with $\bar{f}_{\rm SM}\gamma_\mu f_{\rm SM}$ bilinear results in spin-independent DM-nucleon scattering which is velocity suppressed and thus can be neglected. This type of interaction also provide DM annihilation into SM sector $\left\langle  \sigma v \right\rangle_{kk}$ (specifically with $\bar{f}_{\rm SM}\gamma_\mu \gamma_5 f_{\rm SM}$ bilinear) and dark sector $\left\langle \sigma v \right\rangle_{11\rightarrow 22}$, that are not velocity or momentum suppressed at present.  We now begin with the study of dark matter accumulated inside Sun for multiple dark matter scenarios described above. For this purpose, we obtain the solutions of $N_{k}(t)$ (Eq.~(\ref{2})) with fixed values of $\sigma_{\chi_{k} p},~m_k, f$; $k=1,2$. 
Using the same set of parameter, we then solve for $N^{\prime}_{k}(t)$ (Eq.~(\ref{3})) along with new parameters $\left\langle \sigma v \right\rangle_{kk}$ and $\left\langle \sigma v \right\rangle_{11\rightarrow 22}$.

\begin{figure}
    \begin{center}
        \includegraphics[width=0.45\textwidth]{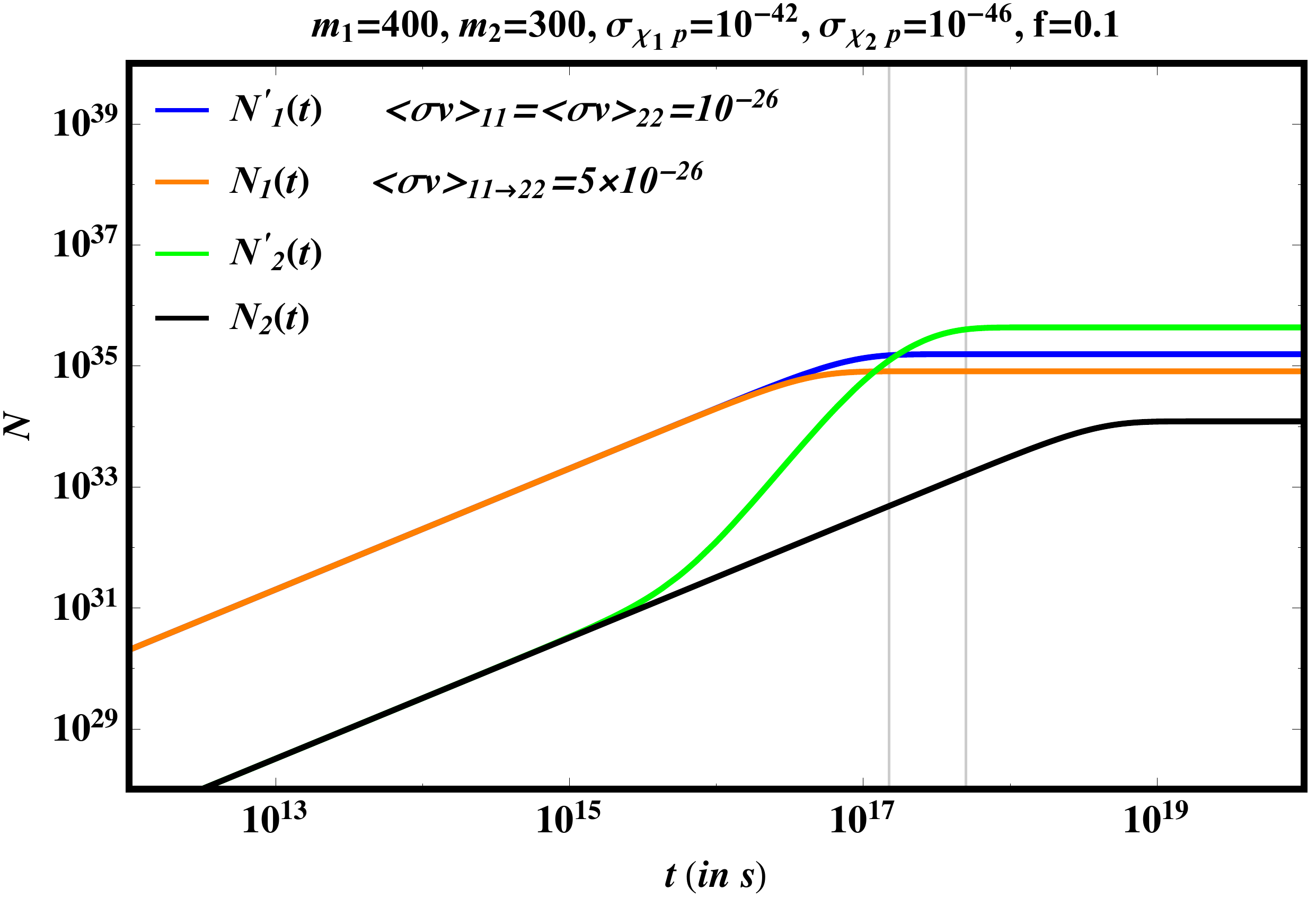}
        \includegraphics[width=0.45\textwidth]{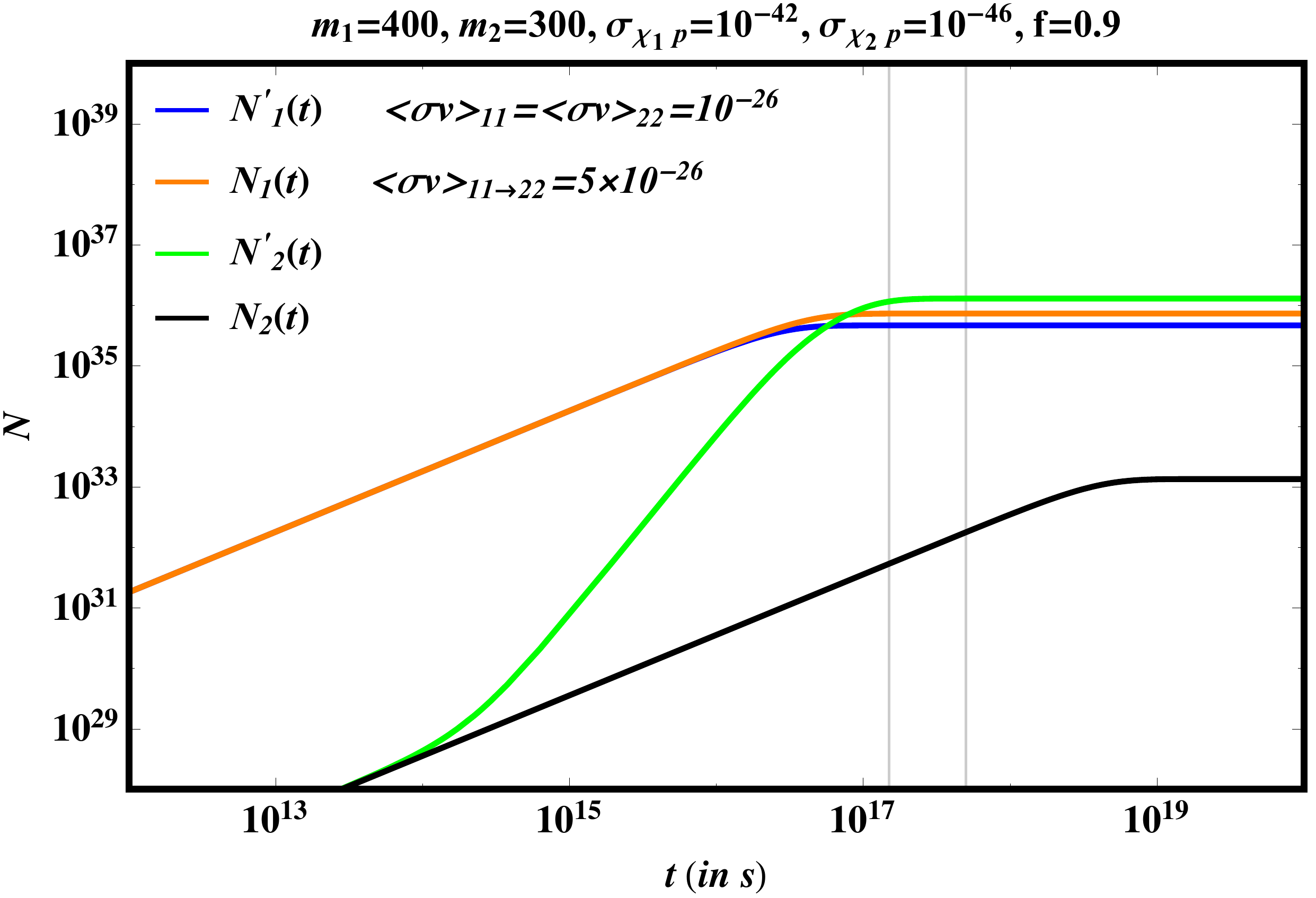}
            \caption{\it Evolution of $\chi_1$ and $\chi_2$ number abundance for $f=0.1$ (upper panel) and $f=0.9$ (lower panel). Dark matter masses $m_k$ are in GeV, $\sigma_{\chi_k p}$ in cm$^2$ and annihilation cross-sections $\left\langle  \sigma v \right\rangle_{kk}$ and $\left\langle \sigma v \right\rangle_{11\rightarrow 22}$ are in cm$^3$ s$^{-1}$ unit.}
        \label{fig:c}
    \end{center}
\end{figure}

\begin{figure}
    \begin{center}
        \includegraphics[width=0.45\textwidth]{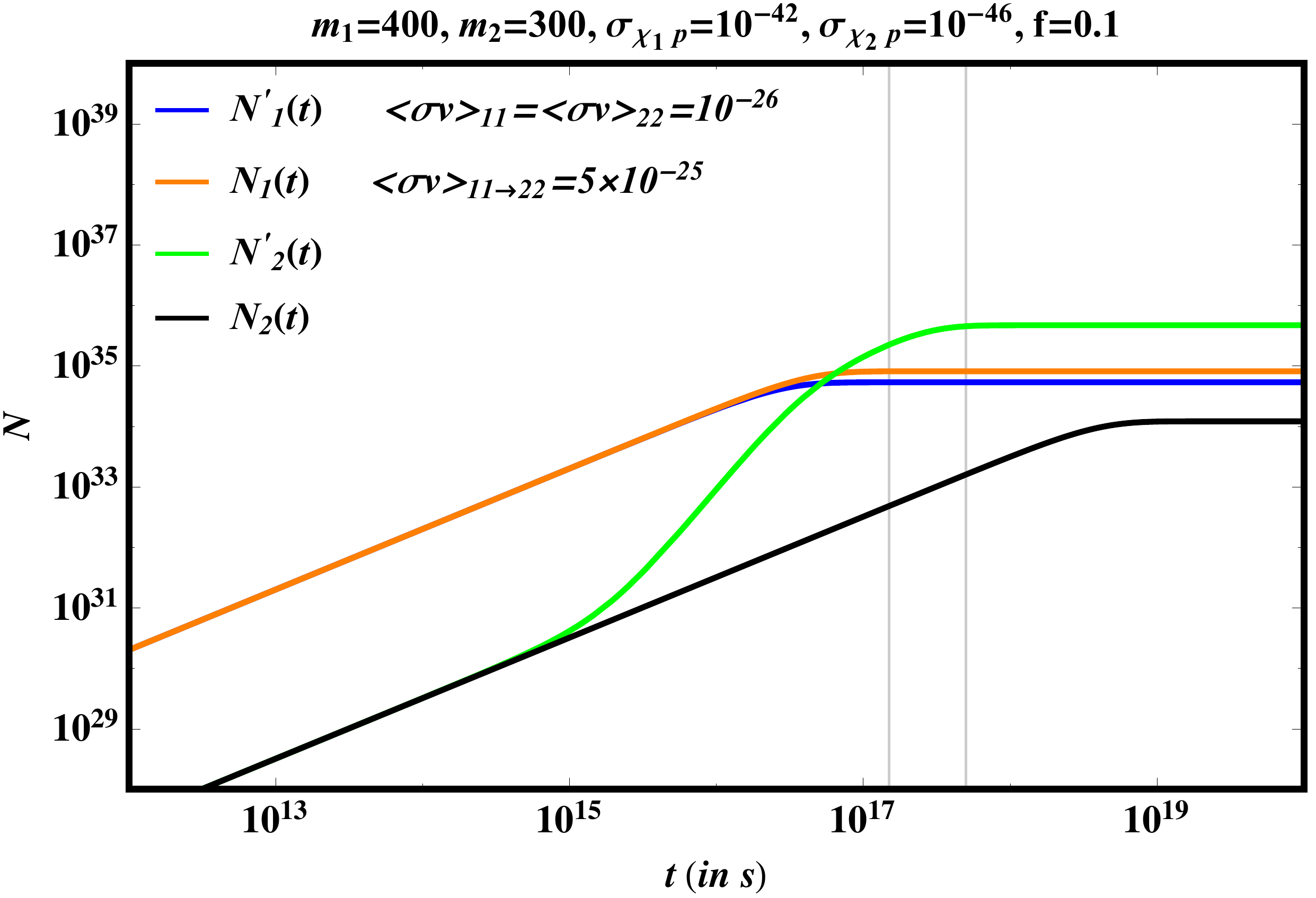}
        \includegraphics[width=0.45\textwidth]{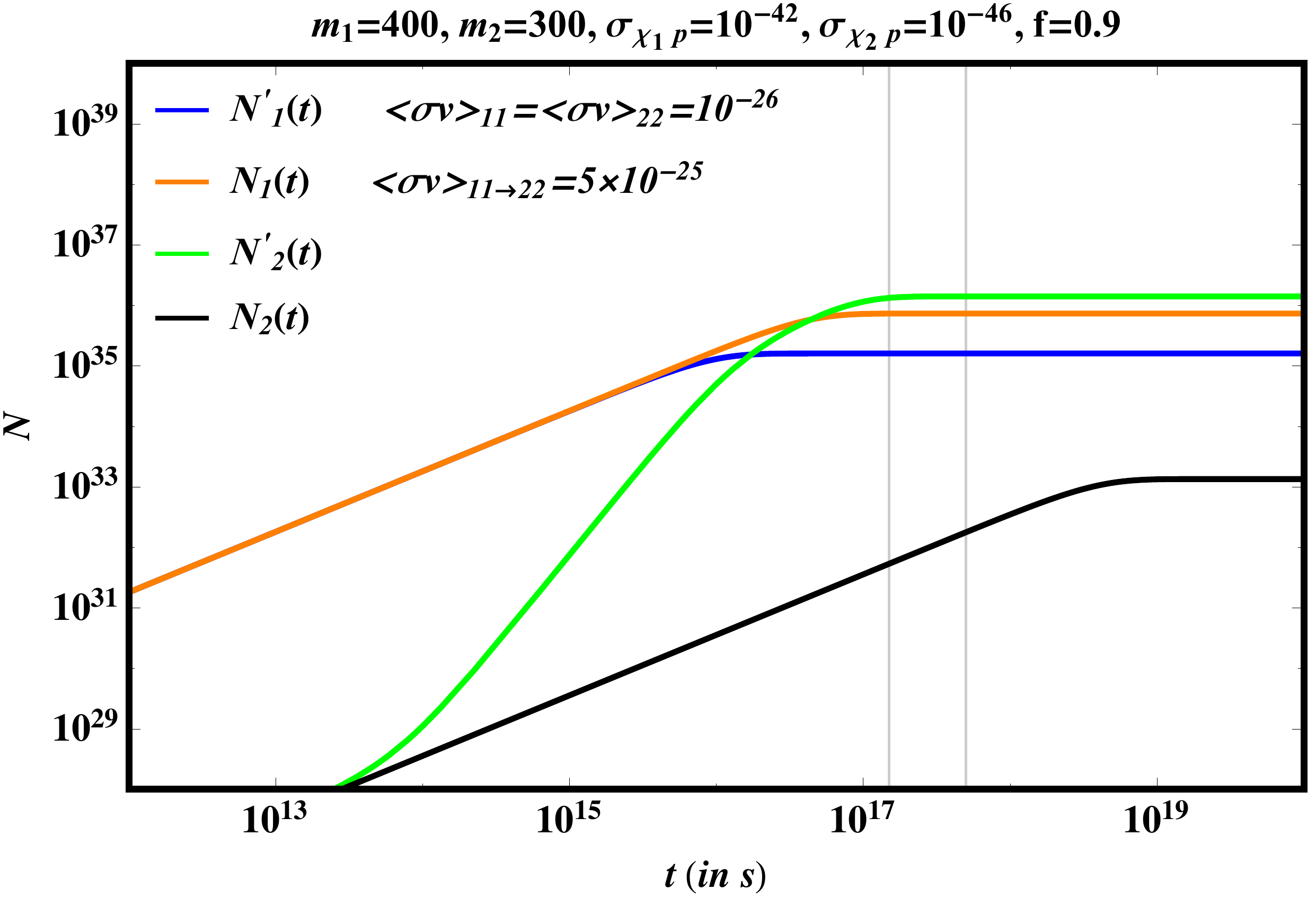}
            \caption{\it Same as Fig.~\ref{fig:c} with $\left\langle \sigma v \right\rangle_{11\rightarrow 22} = 5\times 10^{-25}$ cm$^3$ s$^{-1}$.}
        \label{fig:d}
    \end{center}
\end{figure}

\begin{figure}
    \begin{center}
        \includegraphics[width=0.45\textwidth]{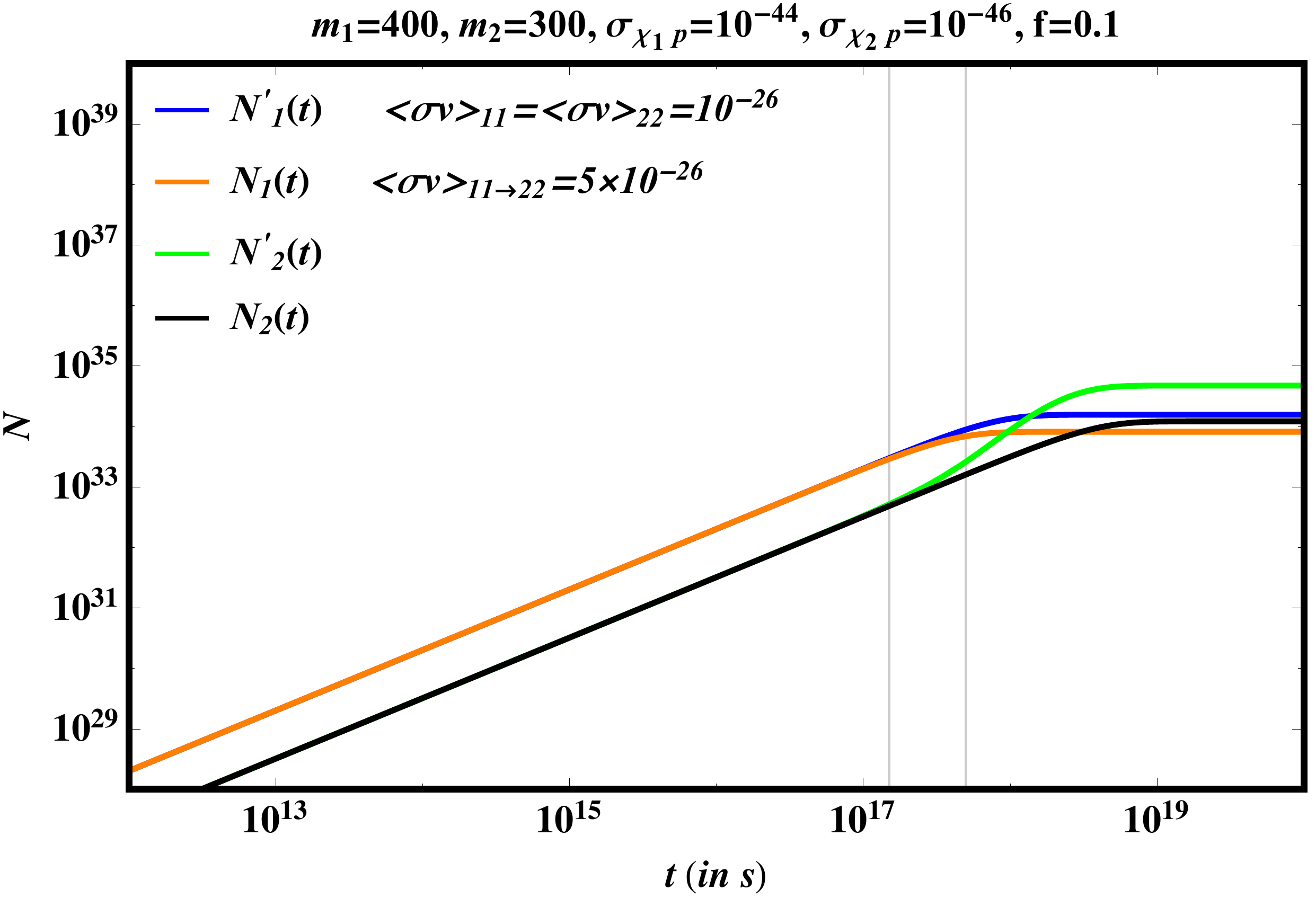}
        \includegraphics[width=0.45\textwidth]{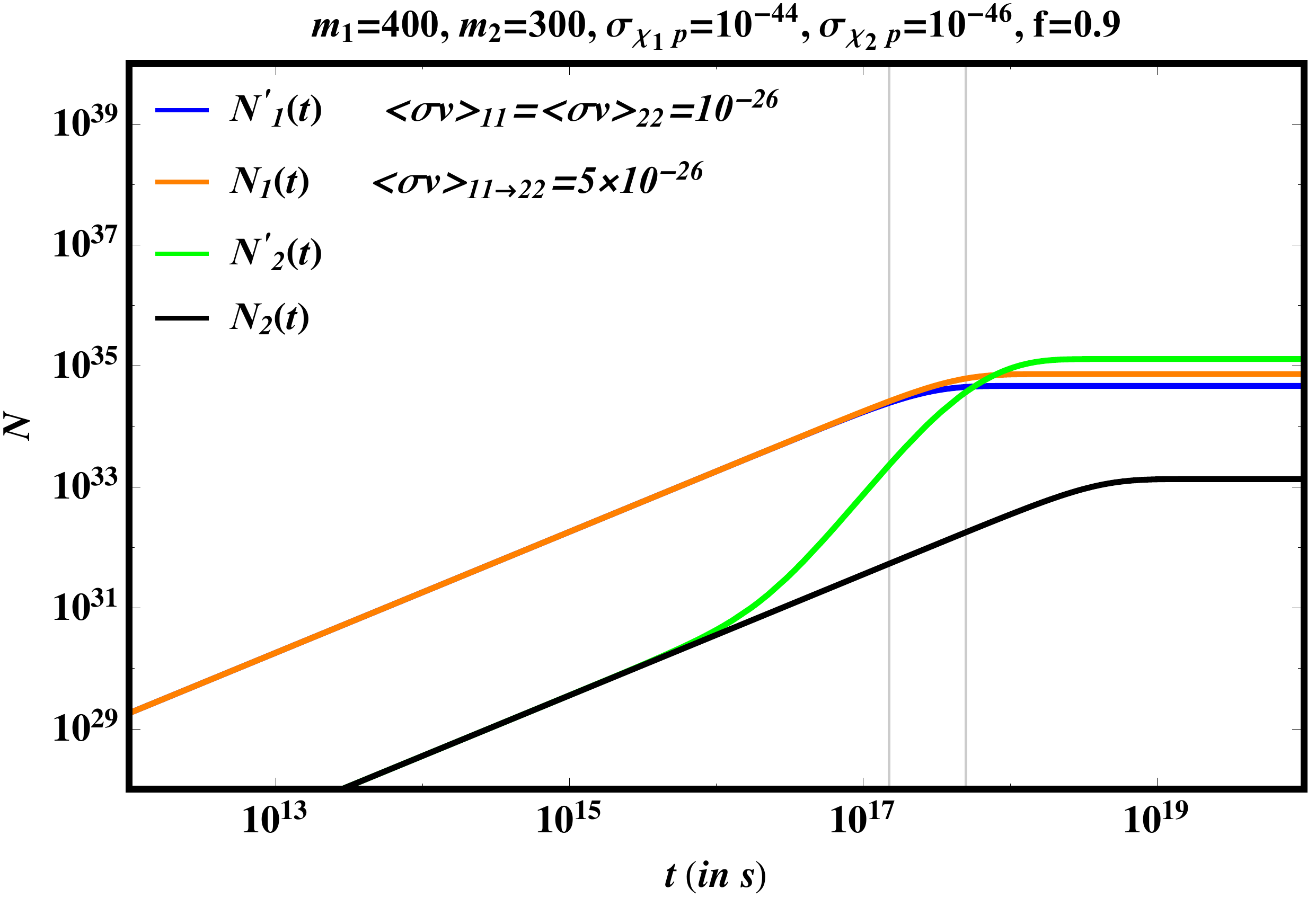}
            \caption{\it Same as Fig.~\ref{fig:c} with $\sigma_{\chi_1 p}=10^{-44}$ cm$^2$.}
        \label{fig:e}
    \end{center}
\end{figure}

To show the number evolution of multi-particle dark matter candidates, some demonstrative plots are presented in Figs.~\ref{fig:c}-\ref{fig:e}.
Abundances of dark matter candidates inside the Sun are obtained by solving Eq.~(\ref{3}) numerically instead of using analytical expressions.
For the purpose of demonstration, following set of parameters are chosen, $m_1=400$ GeV and $m_2=300$ GeV with DM-nucleon scattering cross-section $\sigma_{\chi_{1}p}=10^{-42}$ cm$^2$ and $\sigma_{\chi_{2}p}=10^{-46}$ cm$^2$. 
Using the above set of parameters, dark matter number evolutions are presented in Fig.~\ref{fig:c} for $\left\langle  \sigma v \right\rangle_{kk}=10^{-26}$ cm$^3$ s$^{-1}$, $\left\langle \sigma v \right\rangle_{11\rightarrow 22} = 5\times 10^{-26}$ cm$^3$ s$^{-1}$
using two values of $f=0.1,~0.9$.
It is to be noted that although $\left\langle  \sigma v \right\rangle_{kk};~k=1,2$ is free from the scaling by $f$ as the relation to DM relic abundance is not respected, we use a conservative value well below the upper limits from gamma-ray observation experiments. 
On the other hand, there is no such bound on $\left\langle \sigma v \right\rangle_{11\rightarrow 22}$ from indirect detection experiments.
The number evolution of dark matter candidates in absence of conversion is also presented for comparison obtained from solution of Eq.~(\ref{2}). Vertical lines in plots of Figs.~\ref{fig:c}-\ref{fig:e} denotes the age of Sun and age of Universe respectively to verify if the dark matter $\chi_{1,2}$ will reach equilibrium at the age of Sun. In Fig.~\ref{fig:c}, a large enhancement in the number abundance $N_2^{\prime}$ of dark matter $\chi_2$ is observed with respect to $N_2$ while $N_1^{\prime}$ also deviates from $N_1$ for $\chi_1$. The variation of number abundances are also observed against the change of $f$ as we compare upper and lower panel of Fig.~\ref{fig:c}. Similar plots are depicted in Fig.~\ref{fig:d} with same set of parameters by changing  $\left\langle \sigma v \right\rangle_{11\rightarrow 22} = 5\times 10^{-25}$ cm$^3$ s$^{-1}$
which also reports deviation in number abundance as the conversion effect is more prominent. 
With increase in $f$, the capture rate of dark matter $\chi_2$ reduces resulting smaller number abundance in absence of production from $\chi_1$ (solid black lines) and enhancement in $\chi_1$ abundance (solid red lines).     
The enhancement of $\chi_2$ abundance is related to the excess production from $\chi_1$ dark matter arising from the term $C^{\prime}_{12}  N_1^{\prime 2}$. Increase in $f$ also increases $\chi_2$ production as $C^{\prime}_{12}N_1^{\prime 2}$ term is increased following Eq.~(\ref{N1}), and $\chi_2$ starts deviating from its standard evolution (solid black lines) at some earlier time as seen in Figs.~\ref{fig:c}-\ref{fig:d} and starts to follow the solution of Eq.~(\ref{N2}). 
Note that due to large scattering cross-section $\chi_1$ reaches steady state for both solutions of $N_1$ and $N^{\prime}_1$.
As a result one can obtain equilibrium number abundance for $N_1$ and $N^{\prime}_1$ directly from Eq.~(\ref{eqn:2DM}) and Eq.~(\ref{N1}). Solutions to $N_2$ fails to reach steady state (solid black lines) at $t=t_S$ but in presence of dark matter conversion $\chi_1 \chi_1 \rightarrow \chi_2 \chi_2$, $N_2^{\prime}$ may reach equilibrium (solid green lines) for $t<t_S$,  depending on the values of $f$ and $\left\langle \sigma v \right\rangle_{11\rightarrow 22}$. However, if $\sigma_{\chi_{1}p}$ is small,
then time to reach equilibrium abundances for both the dark matter may exceed the age of Sun. In Fig.~\ref{fig:e} we illustrate such possibilities with $\sigma_{\chi_{1}p}=10^{-44}$ cm$^2$ keeping other parameters fixed as in Fig.~\ref{fig:c}. From Fig.~\ref{fig:e}, we observe although dark matter candidates cannot reach steady state, there could be significant enhancement in the abundance of second dark matter candidate $N_2^{\prime}$ due to large conversion effect.

\begin{figure}
    \begin{center}
        \includegraphics[width=0.45\textwidth]{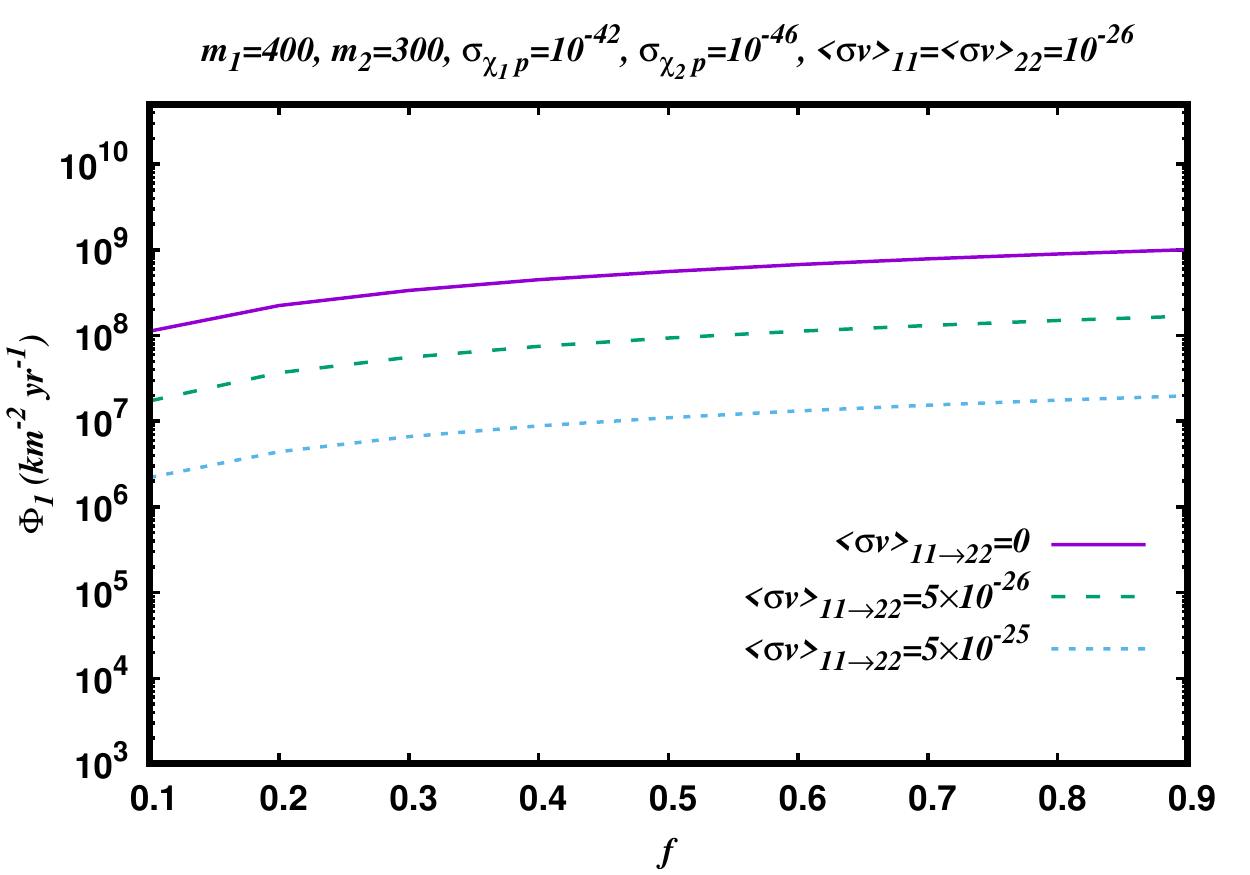}
        \includegraphics[width=0.45\textwidth]{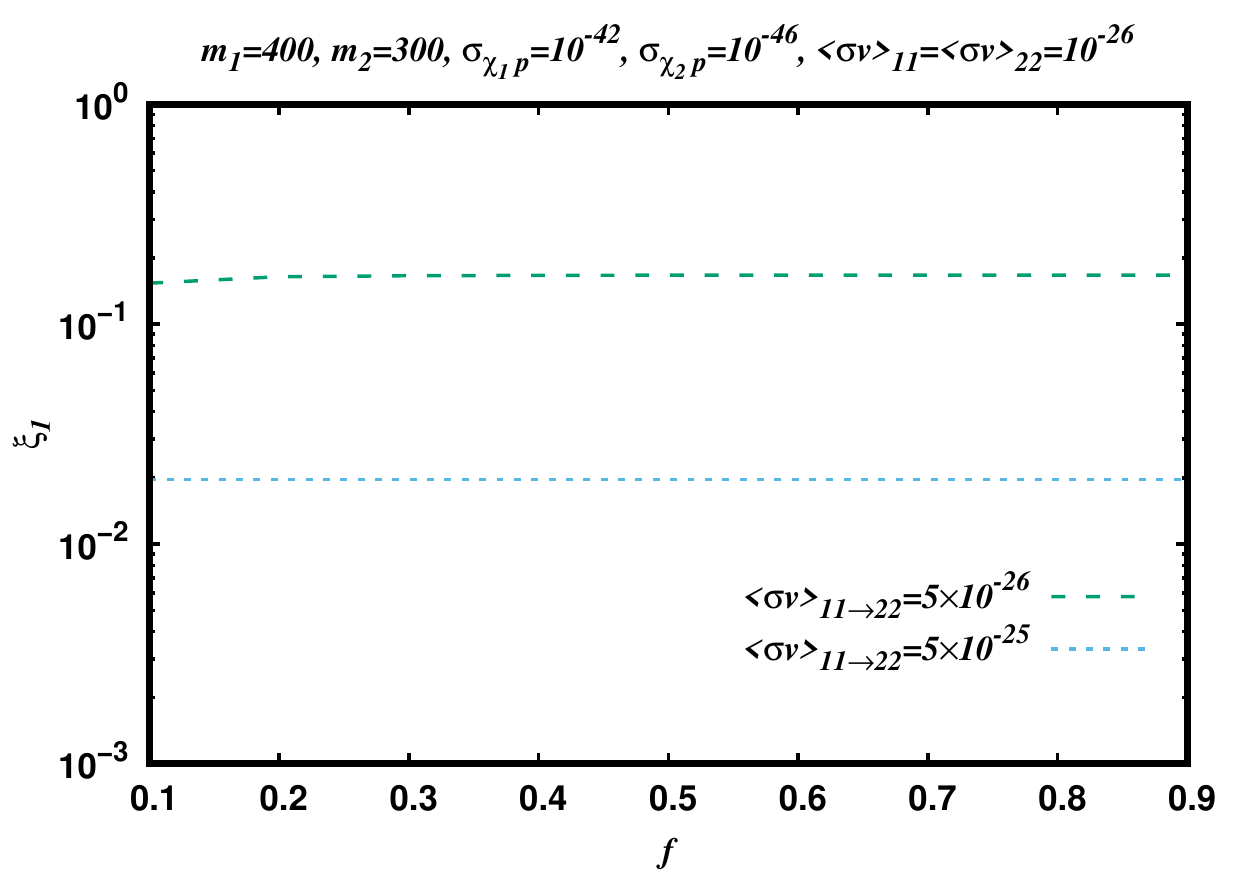}
            \caption{\it Upper panel: $\chi_1$ dark matter annihilation flux vs $f$ for different values of $\left\langle \sigma v \right\rangle_{11\rightarrow 22}$ along with $\left\langle \sigma v \right\rangle_{11\rightarrow 22}=0$ for no conversion. Lower panel: Flux ratio $\xi_1$ against $f$ for chosen set of $\left\langle \sigma v \right\rangle_{11\rightarrow 22}$ values. Units of various quantities $m_k,~\sigma_{\chi_k}, \left\langle \sigma v \right\rangle_{kk}$, and $\left\langle \sigma v \right\rangle_{11\rightarrow 22}$ are same as in Fig.~\ref{fig:c}.}
        \label{fig:f}
    \end{center}
\end{figure}

\begin{figure}
    \begin{center}
        \includegraphics[width=0.45\textwidth]{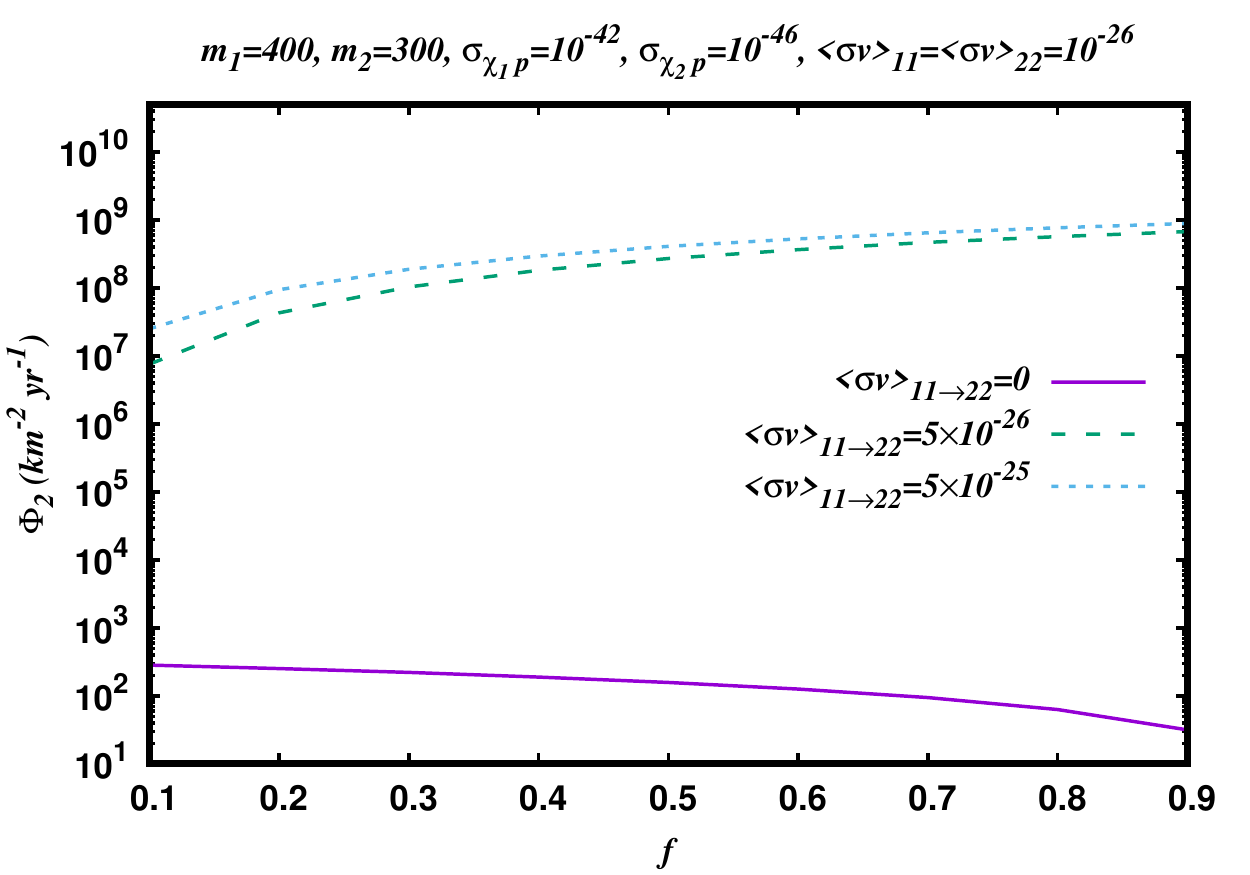}
        \includegraphics[width=0.45\textwidth]{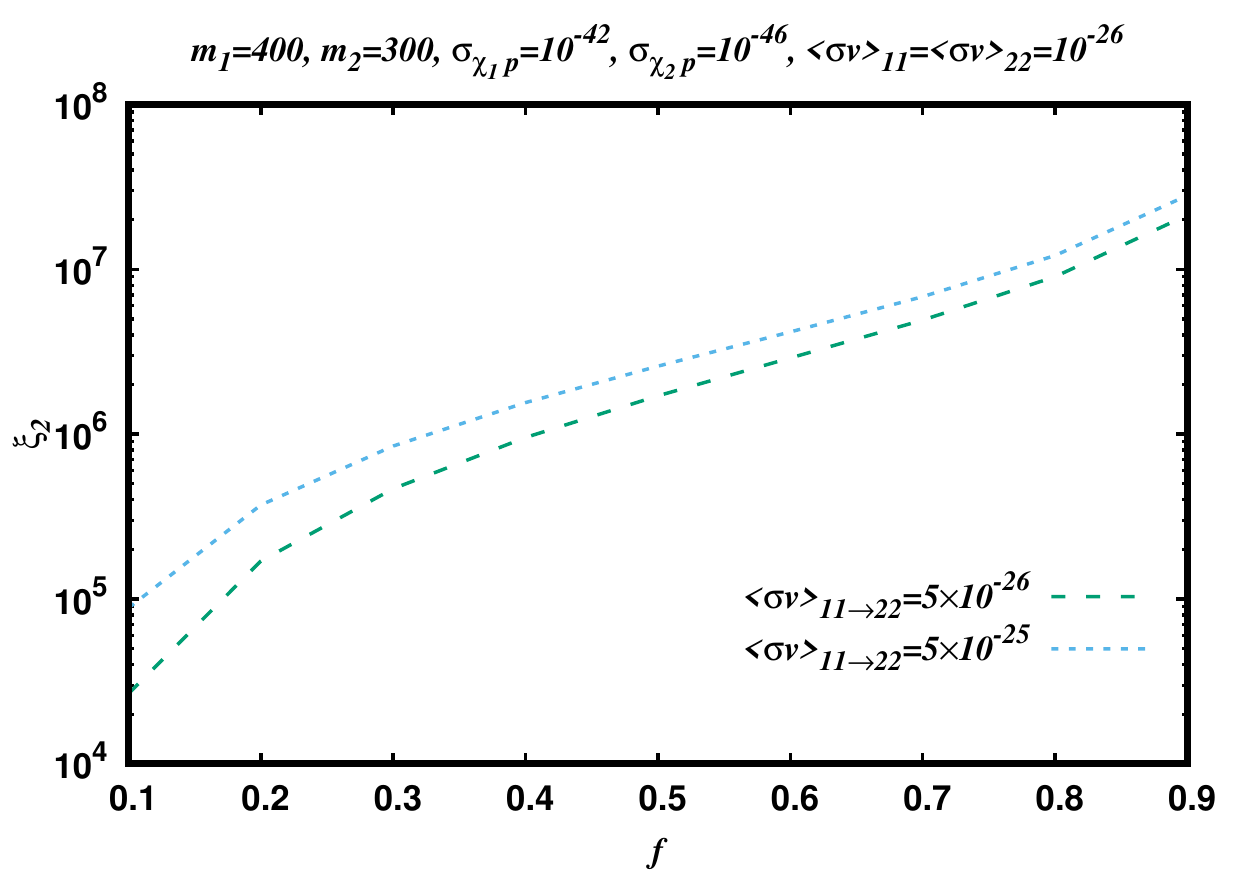}
            \caption{\it Same as Fig.~\ref{fig:f} showing variation of DM annihilation flux with $f$ for different $\left\langle \sigma v \right\rangle_{11\rightarrow 22}$ (upper panel) and efficiency factor $\xi_2$ vs $f$ (lower panel) for dark matter candidate $\chi_2$.}
        \label{fig:g}
    \end{center}
\end{figure}

\begin{figure}
    \begin{center}
        \includegraphics[width=0.45\textwidth]{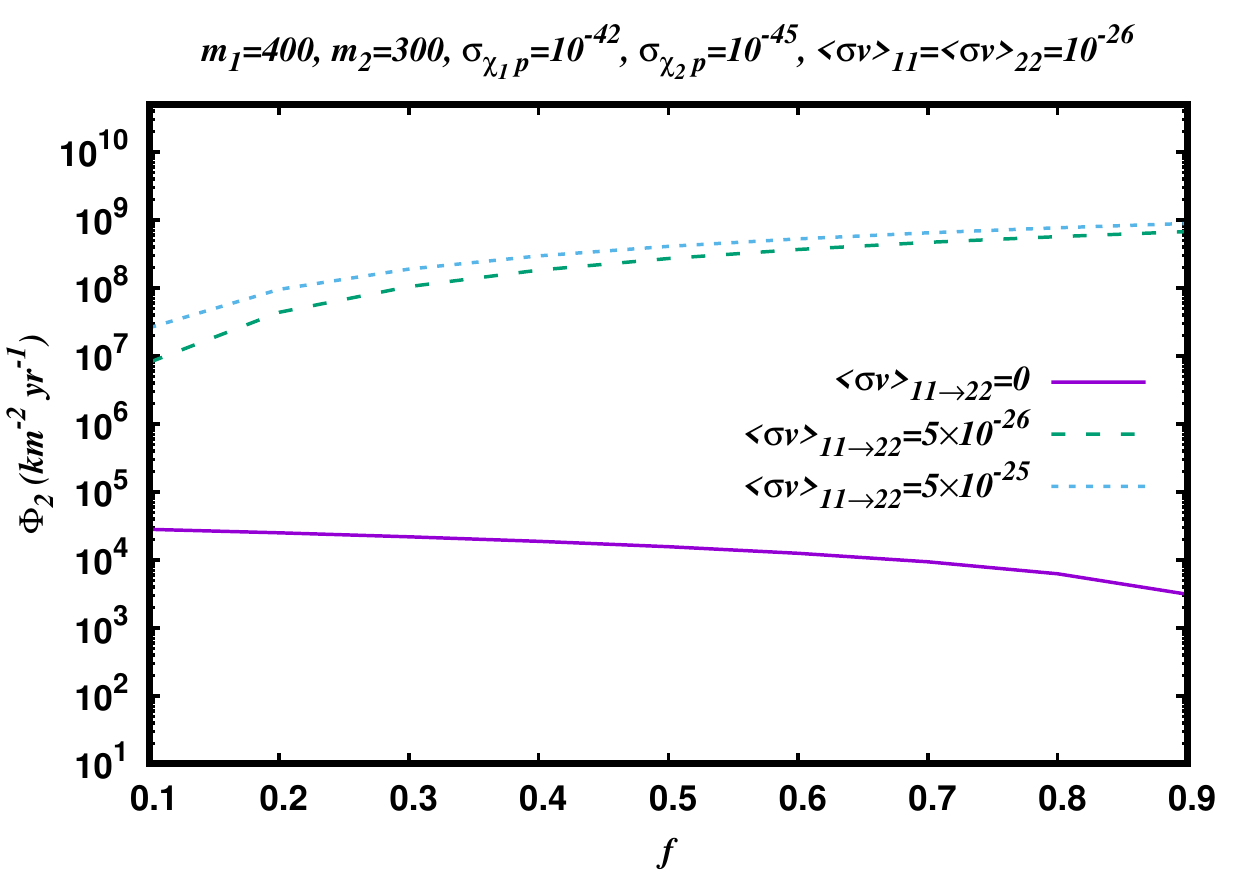}
        \includegraphics[width=0.45\textwidth]{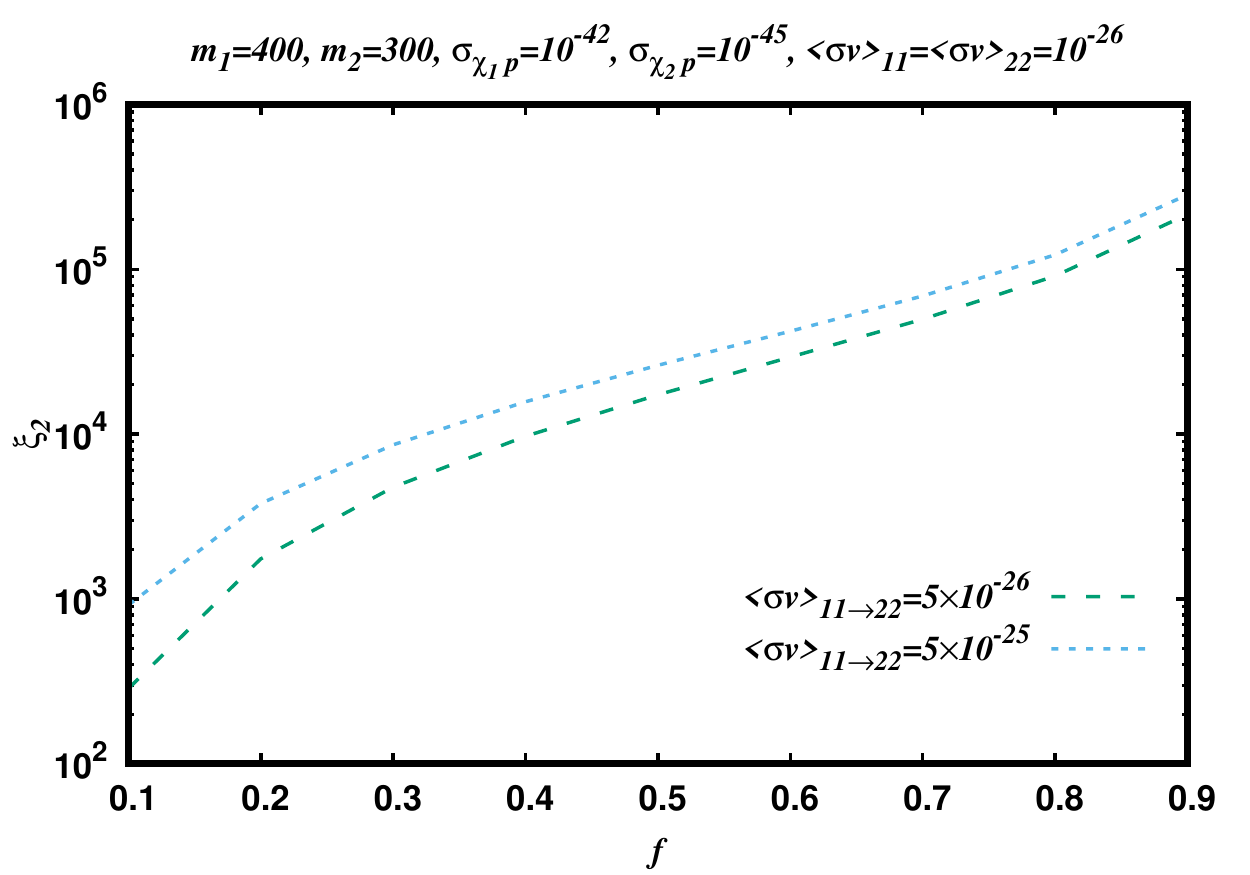}
            \caption{\it Same as Fig.~\ref{fig:g} with $\sigma_{\chi_2 p}=10^{-45}$ cm$^2$.}
        \label{fig:h}
    \end{center}
\end{figure}

Observation made in Figs.~\ref{fig:c}-\ref{fig:e} indicate multi-particle dark matter evolution inside Sun will be significantly different
in presence of DM conversion phenomena. This will also affect DM annihilation flux in the context of multiple dark matter. Let us denote individual DM annihilation flux in absence of $\chi_1 \chi_1 \rightarrow \chi_2 \chi_2$ transfer reaction at $t=t_S$ as
\begin{align}
\Phi_{k} = \frac{\Gamma^{k}_{\text{ann}}}{4\pi D^2}\, , k=1,2, 
\label{flux2}
\end{align}
where annihilation rates of dark matter $\chi_{1,2}$ are expressed as
\begin{align}
\Gamma^1_{\text{ann}}=\frac{C_{1a}}{2f}N^2_1(t_S)\, ,
 \hskip 5mm \Gamma^2_{\text{ann}}=\frac{C_{2a}}{2(1-f)}N^2_2(t_S)\, ,
\label{DMann2} 
\end{align} 
obtained from Eqs.~(\ref{2})-(\ref{eqn:2DM}). Hence one can also obtain the differential neutrino flux following Eq.~(\ref{eq:neutrinoflux})  expressed as

\begin{equation}
\frac{d\Phi^{k}_{\nu_i}}{dE_{\nu_i}}=\Phi_k\left(\frac{dN^k_{\nu_i}}{dE_{\nu_i}}\right)_{x}\, ; k=1,2 
\label{eq:neutrinofluxa}
\end{equation}
where $\left(\frac{dN^k_{\nu_i}}{dE_{\nu_i}}\right)_{x}$ is the spectrum of neutrino and anti-neutrino produced per annihilation
of $k$-th dark matter candidate $\chi_k$ of mass $m_k$ with DM annihilation into final state species $x$. Therefore, corresponding muon flux generated from the neutrino flux is
\begin{eqnarray}
\Phi^{k}_{\mu} &=& \int_{E_\mu^{\rm th}}^{m_k} d E_\mu \int_{E_\mu}^{m_k} d E_{\nu_\mu} \frac{d \Phi^{k}_{\nu_\mu}}{d E_{\nu_\mu}} \\ 
&&\times \left[ \frac{\rho}{m_p} \frac{d \sigma_\nu}{d E_\mu} (E_\mu, E_{\nu_\mu}) R_\mu (E_\mu, E_\mu^{\rm th}) \right] + (\nu \to {\bar \nu})\nonumber \, . 
\label{fluxmu1}
\end{eqnarray}

Similarly, in presence DM transfer ($\chi_1\chi_1 \rightarrow \chi_2\chi_2$), we numerically solve for Eq.~(\ref{3}) and DM annihilation fluxes for both the dark matter candidates are defined as
\begin{align}
\Phi^{\prime}_{k} = \frac{\Gamma^{\prime k}_{\text{ann}}}{4\pi D^2}\, ; k=1,2 \, 
\label{flux2a}
\end{align}
with the modified DM annihilation rates given as
\begin{align}
\Gamma^{\prime k}_{\text{ann}}=\frac{C^{\prime}_{ka}}{2}N^{\prime 2}_k(t_S)\, ; k=1,2 .
\label{DMann2a} 
\end{align} 
Using the modified DM annihilation flux $\Phi^{\prime}_k,~k=1,2$, one can also obtain the muon neutrino flux similar to Eq.~(\ref{fluxmu1})
when dark matter exchange is taken into account

\begin{eqnarray}
\Phi^{\prime k}_{\mu} &=& \int_{E_\mu^{\rm th}}^{m_k} d E_\mu \int_{E_\mu}^{m_k} d E_{\nu_\mu} \frac{d \Phi^{\prime k}_{\nu_\mu}}{d E_{\nu_\mu}} \\ 
&&\times \left[ \frac{\rho}{m_p} \frac{d \sigma_\nu}{d E_\mu} (E_\mu, E_{\nu_\mu}) R_\mu (E_\mu, E_\mu^{\rm th}) \right] + (\nu \to {\bar \nu})\nonumber \, , 
\label{fluxmu2}
\end{eqnarray}
where we use the modified neutrino flux given as $\frac{d\Phi^{\prime k}_{\nu_i}}{dE_{\nu_i}}=\Phi^{\prime}_k\left(\frac{dN^k_{\nu_i}}{dE_{\nu_i}}\right)_{x}\, , k=1,2 $. It is to be noted that the spectrum $\left(\frac{dN^k_{\nu_i}}{dE_{\nu_i}}\right)_{x}$ of neutrino and anti-neutrino generated per dark matter annihilation remains unaltered as it depends on the medium, neutrino oscillation and dark matter mass which determines the energy of produced particles.
Let us now define a quantity $\xi_k=\frac{\Phi^{\prime}_k}{\Phi_k}$ called the efficiency factor. This simply leads to the correlation
\begin{equation}
\frac{d\Phi^{\prime k}_{\nu_i}}{dE_{\nu_i}}=\xi_k \frac{d\Phi^{k}_{\nu_i}}{dE_{\nu_i}}\, ; k=1,2 \\
\label{eq:neutrino_fluxb}
\end{equation}
Eq.~(\ref{eq:neutrino_fluxb}) directly implies $\Phi^{\prime k}_{\mu}=\xi_k \Phi^{k}_{\mu}$ for $k$-th dark matter candidate as $\xi_k$ is a simple scaling with no effect on $\frac{dN^k_{\nu_i}}{dE_{\nu_i}}$ spectrum and has no energy dependence i.e; $\xi_k\neq \xi_k(E_{\mu_\nu}, E_\mu)$. It is to be noted that the efficiency factor $\xi_k$ is a function of many parameters $f,~m_k,~\sigma_{\chi_k p},~\left\langle  \sigma v \right\rangle_{kk},~k=1,2$ and $\left\langle  \sigma v \right\rangle_{11\rightarrow 22}$. Hence, a simple correlation between DM annihilation flux and muon flux obtained, $\xi_k=\frac{\Phi^{\prime}_k}{\Phi_k}=\frac{\Phi^{\prime k}_{\mu}}{\Phi^{k}_{\mu}}$. The parameter $\xi_k$ being a ratio, is also a very intelligent choice to observe the boost or suppression in muon flux obtained from a dark matter candidate $\chi_1$ or $\chi_2$ as it completely avoids the background effects. Therefore, we use the parameter $\xi_k$
as a tool for the study of multi-particle dark matter dynamics inside the Sun, when the exchange $\chi_1 \chi_1 \rightarrow \chi_2 \chi_2$
is taken into account. Since, $\xi_k$ is completely a new parameter originating from multi-particle DM and their conversion, we find it is interesting to observe the effect of $f$ and  $\left\langle  \sigma v \right\rangle_{11\rightarrow 22}$ on $\xi_k$. In Fig.~\ref{fig:f} (upper panel), the variation of dark matter annihilation flux for DM candidate $\chi_1$ is plotted against $f$ for $\left\langle  \sigma v \right\rangle_{11\rightarrow 22}=5\times10^{-26}$ cm$^3$ s$^{-1}$ and $\left\langle  \sigma v \right\rangle_{11\rightarrow 22}=5\times 10^{-25}$ cm$^3$ s$^{-1}$ and compared the results with the case when conversion is absent leading to the solutions of $N_1$ obtained from Eq.~(\ref{eqn:2DM}) keeping other parameters fixed as considered in Fig.~\ref{fig:c}. A significant reduction in the DM annihilation flux is observed when the process 
$\chi_1 \chi_ \rightarrow \chi_2 \chi_2$ is turned on. This finding is very obvious from the solution of Eq.~(\ref{N1}).
We find that large value of $\sigma_{\chi_1 p}$ drives the DM candidate $\chi_1$ to the steady state and thus the corresponding flux ratio $\frac{\phi^{\prime}_1}{\phi_1}$ is found to be
$\xi_1={\frac{fC^{\prime}_{1a}C_{1c}}{(C^{\prime}_{1a}+C^{\prime}_{12})}}\frac{1}{fC_{1c}}=\frac{C^{\prime}_{1a}}{C^{\prime}_{1a}+C^{\prime}_{12}}$, is independent of the factor $f$ using $C^{\prime}_{1c}=C_{1c}$ for fixed $m_1$ and $\sigma_{\chi_1 p}$. In fact, $\xi_1$ turns out to be 
$\frac{\left\langle  \sigma v \right\rangle_{11}}{\left\langle  \sigma v \right\rangle_{11}+\left\langle  \sigma v \right\rangle_{11\rightarrow 22}}$, is directly reflected in the Fig.~\ref{fig:f}. The efficiency factor $\xi_1$, thus further reduces with increased annihilation $\left\langle  \sigma v \right\rangle_{11\rightarrow 22}$ but remains invariant with variation of $f$. 
This result indicates that, the corresponding efficiency factor must satisfy the condition $\xi_1 \le 1$, as shown in the lower panel of Fig.~\ref{fig:f}. However, this may not be the case for the dark matter candidate $\chi_2$, as its number abundance depends on the production from $\chi_1$. In Fig.~\ref{fig:g}, similar plots are generated for dark matter $\chi_2$ using same set of parameters considered in Fig.~\ref{fig:f}. Abundance of $\chi_2$ inside Sun is obtained directly from numerical solution to Eq.~(\ref{3}) instead of using analytical expression. It is observed in Fig.~\ref{fig:g} that
DM annihilation flux gets boosted by an order $\xi_2=10^{4-7}$ and depends on the value of fractional abundance parameter $f$. Efficiency factor $\xi_2$ is also found to increase with increment in $\left\langle  \sigma v \right\rangle_{11\rightarrow 22}$ as shown in Fig.~\ref{fig:g} as it boosts the conversion effect. 
In Fig.~\ref{fig:h}, we repeat the results for Fig.~\ref{fig:g} only changing $\sigma_{\chi_2 p}=10^{-45}$ cm$^2$. Since, the number evolution of $\chi_1$ is decoupled from evolution of $\chi_2$, flux and efficiency factor obtained for $\chi_1$ is found to be same as Fig.~\ref{fig:f}.
On the other hand, due to increased $\sigma_{\chi_2 p}$, $\chi_2$ abundance increases inside the Sun resulting increase in DM annihilation flux in absence of $\chi_1 \chi_1 \rightarrow \chi_2 \chi_2$ process. As $\left\langle  \sigma v \right\rangle_{11\rightarrow 22}$ is turned on, abundance of $\chi_2$
increases further resulting from transfer mechanism and dominated mostly by conversion phenomena. This leads to large enhancement in  annihilation flux of $\chi_2$ dark matter similar to Fig.~\ref{fig:g}. However, the efficiency factor $\xi_2$ in Fig.~\ref{fig:h} gets reduced by two order due to increase in $\sigma_{\chi_2 p}$ which increases initial $\chi_2$ flux in absence of conversion from $\chi_1$ for all choices of $f$. 
Results obtained in Figs.~\ref{fig:g}-\ref{fig:h}, can be explained with an empirical expression of $\xi_2$, which can directly be obtained from Eq.~(\ref{eqn:2DM}), Eq.~(\ref{N2}) and corresponding expressions of flux $\Phi_{2},~\Phi^{\prime}_{2}$ derived earlier, is given as
\begin{align}
& \xi_{2} \simeq \frac{f}{1-f}\frac{C^{\prime}_{12}}{C^{\prime}_{1a}+C^{\prime}_{12}}\frac{C_{1c}}{C_{2c}}\frac{\tanh^2 \left( \frac{t_s}{\tau^{\prime}_{2\odot}} \right)}{\tanh^2 \left( \frac{t_s}{\tau_{2\odot}} \right)} ,\nonumber \\
& = \frac{f}{1-f}(1-\xi_1)\frac{C_{1c}}{C_{2c}}\frac{\tanh^2 \left( \frac{t_s}{\tau^{\prime}_{2\odot}} \right)}{\tanh^2 \left( \frac{t_s}{\tau_{2\odot}} \right)}\, ,
\label{xi2}
\end{align}
where it is assumed that dark matter $\chi_1$ has reached equilibrium inside the Sun, $C^{\prime}_{1c}=C_{1c}$ and $(1-f)C^{\prime}_{2c}<<C^{\prime}_{12}  N_1^{\prime 2}$ for evolution of $\chi_2$ dark matter. It is found that the expression of $\xi_2$ in Eq.~(\ref{xi2}) is a very good approximation to work with as it does not deviate too much from the expected value of $\xi_2$ derived numerically from solution of 
Eq.~(\ref{3}).
The quantity $\xi_2$ obtained from Eq.~(\ref{xi2}) agrees with the plots in Figs.~\ref{fig:g}-\ref{fig:h} when annihilation cross-section $\left\langle  \sigma v \right\rangle_{11\rightarrow 22}=5\times 10^{-25}$ cm$^3$ s$^{-1}$ is taken into account. 
Minor deviation from exact solution to $\chi_2$ abundance may occur as the solution of Eq.~(\ref{N2}) results in a slightly larger than expected $\chi_2$ number abundance with respect to number abundance obtained from Eq.~(\ref{3}). 
Although one finds annihilation flux of $\chi_2$ is enhanced by significant amount, it still remains within the upper limit of DM annihilation flux obtained for SD dark matter in Fig.~\ref{fig:a}. With better sensitivity to DM-nucleon spin-dependent scattering cross-section, neutrino detectors will be able to observe such kind of enhancement in DM annihilation flux (or muon flux) and test the multicomponent dark matter hypothesis in near future.

\section{III. Conclusions}
\label{con}

In this study, we explore a case of multi-particle dark matter dynamics inside the Sun including new effects from hidden sector annihilations.  With new parameters in effect, a significant deviation in outcomes (DM annihilation flux or muon flux at detector) from standard single DM scenario is observed as multi DM formalism is taken into account. A simple underlying particle physics model that allows a feasible multi-particle dynamics with hidden sector annihilation reveals some interesting aspects that are reported in the present work. A brief summary of the finding are mentioned below

\begin{itemize}
\item It is found that dark matter annihilation flux from Sun obtained using XENONnT bound on DM-spin-independent nucleon scattering is almost a million order smaller when compared with the DM annihilation flux from spin-dependent DM-nucleon scattering.

\item Dark matter with spin-independent interaction will fail to reach equilibrium DM number abundance due to very small scattering cross-section which further suppress the DM annihilation flux. However, with present limit on scattering cross-section, spin-dependent DM reaches equilibrium inside Sun.

\item For partial contribution with fractional DM abundance, dark matter annihilation flux will be scaled by the fraction of total DM relic abundance $\Phi^{\prime}=f\Phi$ if conversion between two dark matter candidate is absent. Corresponding neutrino flux and muon flux generated at the detector are also found to be suppressed by factor $f$. Interestingly, it is found that the time to reach steady state is independent of the value $f$.

\item For multi-particle dark matter inside Sun with sufficient annihilation $\left\langle  \sigma v \right\rangle_{11\rightarrow 22}$
modifies the DM annihilation flux considerably. Large internal conversion can indeed change the equilibration time of dark matter candidate. In fact, even a dark matter with very small DM-nucleon scattering cross-section could reach equilibrium number abundance inside the Sun as the time to reach equilibrium gets smaller than the age of Sun.

\item A simple two component DM framework is assumed where the heavier DM candidate annihilates into low mass DM. This changes the annihilation flux of DM candidates significantly and a new parameter $\xi$ (efficiency factor) is introduced.
The efficiency factor $\xi_k;~k=1,2$ determines the scale of enhancement or suppression of DM annihilation flux. 
Corresponding muon flux for the dark matter candidate will also gets boosted or suppressed by the factor $\xi_k$.
A suppression in DM annihilation flux would require detectors to reach greater sensitivity while enhancement of the DM annihilation flux will certainly be interesting to be tested with neutrino detectors. The parameter $\xi_1$ for the heavier dark matter $\chi_1$ that converts into 
lighter DM $\chi_2$, is bounded by condition $\xi_1\leq1$, and for $\chi_1$ at steady state can directly be expressed as 
$\xi_1=\frac{\left\langle  \sigma v \right\rangle_{11}}{\left\langle  \sigma v \right\rangle_{11}+\left\langle  \sigma v \right\rangle_{11\rightarrow 22}}$ whereas $\xi_2\simeq10^{4-7}$ is found to be very large boosting the DM annihilation flux of $\chi_2$. In case where capture of $\chi_2$ (i.e; $\sigma_{\chi_2 p}$) becomes subdominant with respect to the
production of $\chi_2$ from annihilation of $\chi_1$ in dark sector, $\xi_2$ is found to be of the form
$\xi_2\simeq\frac{f}{1-f}(1-\xi_1)\frac{C_{1c}}{C_{2c}}\frac{\tanh^2 \left( \frac{t_s}{\tau^{\prime}_{2\odot}} \right)}{\tanh^2 \left( \frac{t_s}{\tau_{2\odot}} \right)}$ for $t_S\geq \tau^{\prime}_{1\odot}$ ($\chi_1$ in steady state), is a very good estimate to work with without solving for $\chi_2$ abundance numerically.

\item The reported changes in the fluxes of dark matter candidates depends on the particle physics interactions. In case where DM annihilation cross-section is velocity or momentum suppressed, DM candidate can also fail to reach steady state inside the Sun and corresponding DM annihilation flux will be severely reduced to be probed by neutrino detectors.  
\end{itemize}
Their still remains a large prospect in both phenomenological and experimental foreground to establish an understanding of multi-particle
dark matter dynamics and their discovery in stellar astrophysics including other astrophysical objects like dwarf stars or planets. For example,
in case planets like earth, although there is large abundance of heavy nucleus instead of Hydrogen, large evaporation of dark matter happens due to less gravity. In Such conditions, the overall dynamics of multi-particle dark matter will be significantly different from massive stars and will be interesting to explore in future works. Dark matter dynamics can also be different for sub-GeV DM candidates, can be explored in future works. With better sensitivity we expect forthcoming neutrino detectors could not only detect DM signature but also verify multi-particle DM hypothesis in near future.

\vskip 5 mm 

\appendix
\section{APPENDIX A: Boltzmann equations for multicomponent dark matter}
\label{app:appA}

Since the present model deals with two dark matter candidates which also interact with themselves, one needs to solve for the coupled Boltzmann equation. Relic density for each of the dark matter candidate is obtained by solving these coupled equations which are written as
\begin{multline}
  \frac{{\rm d} n_{1}}{{\rm d} t} + 3 {\rm H}n_{1}
  =
  - \langle \sigma {\rm v}\rangle_{11} \,
    \left( n_{1}^{2} - {\bar n}_{1}^{2} \right)
  \\
  - \langle \sigma {\rm v}\rangle_{11 \to 22}
    \left( n_2^{2} - \frac{{\bar n}_{2}^{2}}{{\bar n}_{1}^{2}}n_{1}^2 \right)\, ;
  \label{Boltzmann1}
\end{multline}

\begin{multline}
  \frac{{\rm d} n_2}{{\rm d} t} + 3 {\rm H}n_2
  =
  - \langle \sigma {\rm v}\rangle_{22} \left( n_2^{2} - {\bar n}_{2}^{2} \right)
  \\
  +\langle \sigma {\rm v}\rangle_{11 \to 22}
    \left( n_2^{2} - \frac{{\bar n}_{2}^{2}}{{\bar n}_{1}^{2}}n_{1}^2 \right)\, .
  \label{Boltzmann2}
\end{multline}
where ${\rm{m}_1} > {\rm{m}_2}$. Number densities of dark matter particles $\chi_1$ and $\chi_2$, are denoted by $n_k,~k=1,2$ whereas their equilibrium number densities at temperature $T$ are given by ${\bar n}_{k}$. Thermally averaged DM annihilation cross-section into Standard Model sector is given by  $\langle \sigma {\rm v}\rangle_{kk},~k=1,2$ while the same for annihilation cross-section into hidden sector (for $\chi_1 \chi_1 \rightarrow \chi_2 \chi_2$ process) is expressed as $\langle \sigma {\rm v}\rangle_{11 \to 22}$. Solving for coupled Boltzmann equations 
Eqs.~(\ref{Boltzmann1})-(\ref{Boltzmann2}), one obtains relic abundance of DM candidates

\begin{gather}
  \Omega_{k} h^2=2.755\times10^8\frac{m_k}{GeV} \, Y_k(T_0) \, ,
   \quad k=1,2\, ,
  \label{relic}
\end{gather}
where $Y_k=n_k/{\rm s}$, is the comoving number density of $\chi_1$ and $\chi_2$ at present temperature $T=T_0$ with ${\rm s}$ being the entropy density of the Universe.

\vskip 5 mm

\noindent {\bf Acknowledgments} : ADB acknowledges financial support from DST, India, under grant number IFA20-PH250 (INSPIRE Faculty Award). ADB thanks B. Barman for help with useful resources. 
\bibliographystyle{apsrev}
\bibliography{reference}


\end{document}